\documentclass[twocolumn]{aastex631}

\pdfoutput=1 
\usepackage{amsmath,amstext}
\usepackage[T1]{fontenc}
\usepackage{apjfonts} 
\usepackage[figure,figure*]{hypcap}
\usepackage{hyperref}
\usepackage[utf8]{inputenc}
\newcommand{\p}[1]{{\color{cyan}{#1}}}
\renewcommand{\d}[1]{{\color{red}{#1}}}

\newcommand{\angstrom}{\mbox{\normalfont\AA}}
\newcommand{\g}[1]{{\color{cyan}{#1}}}

\newcommand{\f}[2]{{\ensuremath{\mathchoice%
        {\dfrac{#1}{#2}}
        {\dfrac{#1}{#2}}
        {\frac{#1}{#2}}
        {\frac{#1}{#2}}
        }}}

\renewcommand{\vec}[1]{ \textit{\textbf{#1}} }

\newcommand{\degree}{\ensuremath{^\circ}}

\newcommand{\kms}{km s$^{-1}$}
\newcommand{\Rsun}{R_\odot}
\shorttitle{Orbiter S-Web}
\shortauthors{Baker, D. et al.}

\begin{document}

\title{Observational Evidence of S-Web Source of the Slow Solar Wind}
\author{Baker, D.}
\affiliation{University College London, Mullard Space Science Laboratory, Holmbury St. Mary, Dorking, Surrey, RH5 6NT, UK}

\author[0000-0001-8215-6532]{D\'emoulin, P.}
\affiliation{LESIA, Observatoire de Paris, Universit\'e PSL, CNRS, Sorbonne Universit\'e, Univ. Paris Diderot, Sorbonne Paris Cit\'e, 5 place Jules Janssen, 92195 Meudon, France}
\affiliation{Laboratoire Cogitamus, rue Descartes, 75005 Paris, France}

\author[0000-0003-2802-4381]{Yardley, S. L.}
\affiliation{Department of Meteorology, University of Reading, Reading, UK}
\affiliation{University College London, Mullard Space Science Laboratory, Holmbury St. Mary, Dorking, Surrey, RH5 6NT, UK}
\affiliation{Donostia International Physics Center (DIPC), Paseo Manuel de Lardizabal 4, 20018 San Sebastián, Spain}

\author{Mihailescu, T.}
\affiliation{University College London, Mullard Space Science Laboratory, Holmbury St. Mary, Dorking, Surrey, RH5 6NT, UK}

\author{van Driel-Gesztelyi, L.}
\affiliation{University College London, Mullard Space Science Laboratory, Holmbury St. Mary, Dorking, Surrey, RH5 6NT, UK}
\affiliation{LESIA, Observatoire de Paris, Universit\'e PSL, CNRS, Sorbonne Universit\'e, Univ. Paris Diderot, Sorbonne Paris Cit\'e, 5 place Jules Janssen, 92195 Meudon, France}
\affiliation{Konkoly Observatory, Research Centre for Astronomy and Earth Sciences, Konkoly Thege \'ut 15-17., H-1121, Budapest, Hungary}

\author{D'Amicis, R.}
\affiliation{National Institute for Astrophysics, Institute for Space Astrophysics and Planetology, Rome, Italy}

\author{Long, D. M.}
\affiliation{University College London, Mullard Space Science Laboratory, Holmbury St. Mary, Dorking, Surrey, RH5 6NT, UK}
\affiliation{Astrophysics Research Centre, School of Mathematics and Physics, Queen’s University Belfast, University Road, Belfast, BT7 1NN, Northern Ireland, UK}

\author{To, A. S H.}
\affiliation{University College London, Mullard Space Science Laboratory, Holmbury St. Mary, Dorking, Surrey, RH5 6NT, UK}

\author{Owen, C. J.}
\affiliation{University College London, Mullard Space Science Laboratory, Holmbury St. Mary, Dorking, Surrey, RH5 6NT, UK}

\author{Horbury, T. S.}
\affiliation{Imperial College London, Blackett Laboratory, South Kensington, SW7 2AZ}

\author{Brooks, D. H.}
\affiliation{College of Science, George Mason University, 4400 University Drive, Fairfax, VA 22030, USA}

\author{Perrone, D.}
\affiliation{ASI -- Italian Space Agency, Via del Politecnico, s.n.c 00133 – Roma, Italia}

\author[0000-0001-9726-0738]{French, R. J.}
\affiliation{National Solar Observatory, 3665 Innovation Drive, Boulder CO 80303}
\author{James, A. W.}
\affiliation{University College London, Mullard Space Science Laboratory, Holmbury St. Mary, Dorking, Surrey, RH5 6NT, UK}
\affiliation{European Space Agency (ESA), European Space Astronomy Centre (ESAC), Camino Bajo del Castillo, s/n, 28692 Villanueva de la Ca\~{n}ada, Madrid, Spain}

\author{Janvier, M.}
\affiliation{Universit\'e Paris-Saclay, CNRS, Institut d'Astrophysique Spatiale, 91405 Orsay, France}
\affiliation{Laboratoire Cogitamus, rue Descartes, 75005 Paris, France}

\author{Matthews, S.}
\affiliation{University College London, Mullard Space Science Laboratory, Holmbury St. Mary, Dorking, Surrey, RH5 6NT, UK}

\author{Stangalini, M.}
\affiliation{ASI -- Italian Space Agency, Via del Politecnico, s.n.c 00133 – Roma, Italia}

\author{Valori, G.}
\affiliation{Max Planck Institute for Solar System Research, Justus-von-Liebig-Weg 3, D-37077 G\"ottingen, Germany}
\author{Smith, P.}
\affiliation{University College London, Mullard Space Science Laboratory, Holmbury St. Mary, Dorking, Surrey, RH5 6NT, UK}
\author{Aznar Cuadrado, R.}
\affiliation{Max Planck Institute for Solar System Research, Justus-von-Liebig-Weg 3, D-37077 G\"ottingen, Germany}

\author{Peter, H.}
\affiliation{Max Planck Institute for Solar System Research, Justus-von-Liebig-Weg 3, D-37077 G\"ottingen, Germany}

\author{Schuehle , U.}
\affiliation{Max Planck Institute for Solar System Research, Justus-von-Liebig-Weg 3, D-37077 G\"ottingen, Germany}

\author{Harra, L.}
\affiliation{PMOD/WRC, Dorfstrasse 33 7260 Davos Dorf, Switzerland}
\affiliation{ETH-Z\"{u}rich, H\"{o}nggerberg campus, HIT building, Z\"{u}rich, Switzerland}
\author{Barczynski, K.}
\affiliation{PMOD/WRC, Dorfstrasse 33 7260 Davos Dorf, Switzerland}
\affiliation{ETH-Z\"{u}rich, H\"{o}nggerberg campus, HIT building, Z\"{u}rich, Switzerland}

\author{Berghmans, D.}
\affiliation{Solar-Terrestrial Centre of Excellence – SIDC, Royal Observatory of Belgium, Ringlaan -3- Av. Circulaire, 1180 Brussels, Belgium}
\author{Zhukov, A. N.}
\affiliation{Solar-Terrestrial Centre of Excellence – SIDC, Royal Observatory of Belgium, Ringlaan -3- Av. Circulaire, 1180 Brussels, Belgium}
\affiliation{Skobeltsyn Institute of Nuclear Physics, Moscow State University, 119992 Moscow, Russia}
\author{Rodriguez, L.}
\affiliation{Solar-Terrestrial Centre of Excellence – SIDC, Royal Observatory of Belgium, Ringlaan -3- Av. Circulaire, 1180 Brussels, Belgium}
\author{Verbeeck, C.}
\affiliation{Solar-Terrestrial Centre of Excellence – SIDC, Royal Observatory of Belgium, Ringlaan -3- Av. Circulaire, 1180 Brussels, Belgium}

\begin{abstract}
From 2022 March 18-21, active region (AR) 12967 was tracked simultaneously by Solar Orbiter (SO) at 0.35 au and Hinode/EIS at Earth.
During this period, strong blue-shifted plasma upflows were observed along a thin, dark corridor of open magnetic field originating at the AR's leading polarity and continuing towards the southern extension of the northern polar coronal hole.
A potential field source surface (PFSS) model shows large lateral expansion of the open magnetic field along the corridor. Squashing factor 
$Q$-maps of the large scale topology further confirm super-radial expansion in support of the S-Web theory for the slow wind.
The thin corridor of upflows is identified as the source region of a slow solar wind stream characterized by $\sim 300$ \kms\ 
velocities, low proton temperatures of $\sim$5 eV, extremely high density $>100$ cm$^{-3}$, and a short interval of moderate Alfv\'enicity accompanied by  switchback events. 
When the connectivity changes from the corridor to the eastern side of the AR, the in situ plasma parameters of the slow solar wind indicate a distinctly different source region.
These observations provide strong evidence that the narrow open field corridors, forming part of the S-web, produce some extreme properties in their associated solar wind streams.

\end{abstract}



\section{Introduction}
\label{sect_Introduction} 
The solar corona continuously expands outward as hot, magnetized streams of plasma that fill the heliosphere.  
These streams are generally classified as the fast and slow solar wind (SW) based on a velocity threshold of 500 \kms\ 
 \citep[e.g.][]{temmer21}.  
The fast SW is quasi-uniform, exhibiting smooth, slow changes in plasma parameters, whereas the slow SW tends to be highly variable, especially in plasma composition and density.
The contrasting characteristics of the two SW streams reflect their differing source regions.
Fast SW originates from coronal holes (CHs); the plasma travels unadulterated along open field from the chromosphere/corona into the heliosphere \citep[e.g.][]{zirker77,cranmer09}.
However, the sources of the slow SW are found in the closed field regions on the Sun, e.g. active regions, coronal hole boundaries (CHB), streamers \citep[e.g.][]{geiss95,mccomas98,brooks15,abbo16}.
The highly variable nature of the plasma suggests that the slow SW is released onto open field lines by stochastic processes such as magnetic reconnection.  
How plasma confined to the closed field of the Sun contributes to the solar wind remains an open question. 
Results from recent missions, such as ESA's Solar Orbiter \citep[SO;][]{muller20,garcia21} and NASA's Parker Solar Probe \citep[PSP;][]{fox16}, which are parts of the Heliosphysics System Observatory, are providing new insights into the physical processes and environment where the slow SW is formed and accelerated.

The S-web model proposed by \cite{antiochos11} provides a theoretical framework for the origin of the slow SW.
According to this model, the corona is filled with a network of narrow open field corridors bounded by a web of separatrix surfaces or quasi-separatrix layers \cite[QSLs,][]{demoulin96}. These boundaries are defined by  discontinuous/drastic changes of the field line mapping from the photosphere to the source surface.
The narrow, sometimes infinitesimal, corridors are topologically robust features that link coronal holes on the Sun \citep{antiochos07,antiochos11}.  
The topology of the S-web consists of giant arcs of open--closed magnetic flux boundary layers extending 10s of degrees in longitude  \citep[][]{higginson17b,scott18}. 

Interchange reconnection is proposed as the mechanism responsible for the release of the slow SW at the S-web arcs.
Three-dimensional magnetohydrodynamic (MHD) simulations of \citet{higginson17a} demonstrated that photospheric convective motions on supergranular scales drive the magnetic field differently within open/closed field regions. 
This results in ongoing dynamics along the S-web. 
Multiple interchange reconnection events are induced there and plasma is exchanged between the open and closed flux domains.
Interchange reconnection is also induced in pseudo-streamer configurations at their null points and separators \citep{aslanyan21}.
Recently, \citet{chitta22} presented Extreme UltraViolet (EUV) and white-light observations of the highly structured mid-corona that show indications of the interchange reconnection dynamics consistent with the MHD simulations of the S-web.

In this work, we provide observational evidence of an S-web source region and its associated slow SW stream.
Plasma parameters are determined within the narrow corridor, which is an open field source region, from spectroscopic observations acquired by Hinode's EUV Imaging Spectrometer \citep[EIS;][]{culhane07}. 
The associated slow SW stream is characterized from in situ measurements made by SO's Magnetometer \citep[MAG;][]{horbury20} and Solar Wind Analyser Proton and Alpha particle Sensor  \citep[SWA-PAS;][]{owen20}.
SO's Extreme Ultraviolet Imager (EUI)/Full Sun Imager \citep[FSI;][]{rochus20} coupled with magnetic field observations and modeling are employed to connect the SW to its probable source region.

\begin{figure*}[t]
\epsscale{1.1}
\plotone{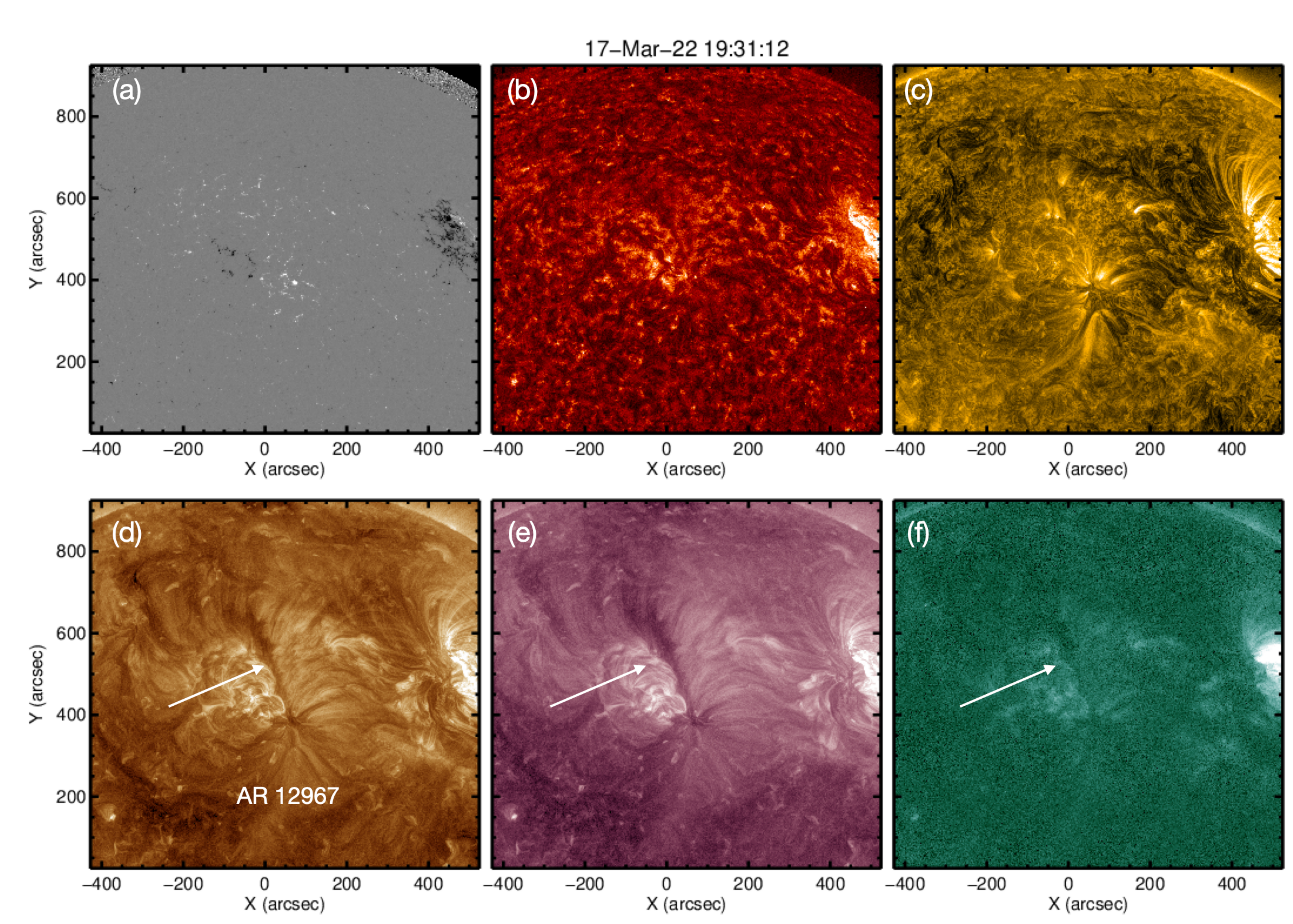}
\caption{SDO/HMI line of sight magnetogram (a) and SDO/AIA 304 $\angstrom$ (b), 171 $\angstrom$ (c), 193 $\angstrom$ (d), 211 $\angstrom$ (e), and 94 $\angstrom$ (f) images of AR 12967 at 2022 March 17 19:31 UT.
Arrow indicates the dark open field corridor described in the text (see Section \ref{sec:aia}).
The figure has an included movie labeled 20220314$\_$multi$\_$panel.mp4.
It shows the chromospheric/coronal evolution of AR 12967 from 2022 March 14 12:00 UT to 2022 March 20 23:56 UT (duration is 57 s).
\label{fig:multi}} 
\end{figure*}


\section{Solar Orbiter Target Region}\label{SO_stuff}

In March 2022, SO was at perihelion when the science observation phase of the mission commenced. 
During this time, a number of the Solar Orbiter Observing Plans (SOOPs) selected by the high level Science Working Team (SWT) were run.
SO's complex payload comprising 6 remote sensing and 4 in situ instruments requires coordinated planning in order to ensure that the mission's scientific objectives are met.
SOOPs are the primary mechanism for planning and coordinating SO observations among its own instruments as well as with other ground-based and space-borne assets. 
SOOPs are especially important for observations acquired during the three 10-day remote sensing windows scheduled for each orbit \citep[for a complete account of SO's planning, see][]{zouganelis20}.
The L$\_$SMALL$\_$HRES$\_$HCAD$\_$Slow-Wind-Connection SOOP was selected to be among the first to be run.
Its stated aim is to identify the sources of the SW and it 
was designed to address in part one of the four top level science questions of the mission: "What drives the solar wind and where does the coronal magnetic field originate?" \citep{muller20}.

The ideal target source region of the SOOP is at the boundary/interface of open and closed magnetic field either at the edges of active regions close to low-latitude open field or at CHBs.
Open field can provide pathways for plasma to travel from the source region to SO where it is detected in situ.
Remote sensing instruments are pointed at the target with the expectation that the spacecraft crosses the boundary/interface a number of days later, depending on a number of factors including SO's distance from the Sun and the SW velocity.
Connectivity of the slow wind stream is then confirmed by matching plasma parameters measured at the Sun and at the SO spacecraft in combination with magnetic field modeling.
AR 12967 provided the perfect slow wind source region target for SO when the spacecraft was $\sim$0.35 au from the Sun.
Utilizing spectroscopic measurements, magnetic topology, and in situ observations, AR 12967 is identified as being the likely source region magnetically connected to the slow wind stream detected in situ (see Section \ref{sec:connect}).
A complete account of the target selection, connectivity, and science observations taken when SO was at/near to perihelion is provided in \cite{yardley23}.

\section{Remote Sensing Observations}\label{sec:obs}
\subsection{SDO/HMI Observations}\label{sec:hmi}

AR 12967 rotated onto the solar disk on 2022 March 12 (from the viewpoint of Earth).
Its leading polarity contained a small, coherent positive sunspot bordered to the north and south by dispersed positive field. 
This configuration is shown in the magnetogram from the Helioseismic and Magnetic Imager  \citep[HMI;][]{scherrer12} on board the Solar Dynamics Observatory \citep[SDO; ][]{pesnell12} in Figure \ref{fig:multi}a. 
By 2022 March 18, the sunspot was no longer visible in the HMI magnetograms.
The following negative polarity was fully dispersed from the time it was first observed at the east limb, suggesting that the AR was already in its decay phase.
At central meridian passage, the AR's total unsigned magnetic flux was $\sim$3.5$\times$10$^{21}$ Mx, typical for a medium-sized AR with a lifetime measured in weeks rather than months \citep{lvdg15}.

The distribution of the photospheric magnetic field remained broadly unchanged throughout the AR's transit across the solar disk (see the included movie labeled 20220314$\_$multi$\_$panel.mp4).
Its negative following polarity field was embedded in the surrounding positive field in the southern section of the AR.
In the AR's northern section, the dispersed positive field extended toward the northern polar CH. 

\subsection{SDO/AIA Observations}\label{sec:aia}

Images from the 304 $\angstrom$ (log $T$ = 4.7), 171 $\angstrom$ (log $T$ = 5.8), 193 $\angstrom$ (log $T$ = 6.2 and 7.3), 211 $\angstrom$ (log $T$ = 6.3), and 94 $\angstrom$ (log $T$ = 6.8) channels of SDO's Atmospheric Imaging Assembly \citep[AIA;][]{lemen12} are displayed in Figure \ref{fig:multi}b--f.
The multi-panel image is taken from the movie 20220314$\_$multi$\_$panel.mp4.
The movie shows the chromospheric/coronal evolution of AR 12967 from 2022 March 14 12:00 UT to 2022 March 20 23:56 UT.
Each image has a field-of-view (FOV) of 900$\arcsec\times850\arcsec$ encompassing AR 12967 and the surrounding dispersed magnetic field.
Bright core loops connecting the AR's main polarities are visible in the hotter AIA channels. 
Less bright, longer loops extend from around the positive sunspot to the negative fields of the nearby AR 12965 in the west and quiet Sun in the south.
Loops rooted along the channel of positive field in the north appear to connect with the negative polarity field of AR 12967 on the eastern edge and negative field of AR 12965 on the western edge, forming a dark, narrow corridor that extends from the positive sunspot toward the north polar CH.

\begin{figure*}[t]
\epsscale{1.2}
\plotone{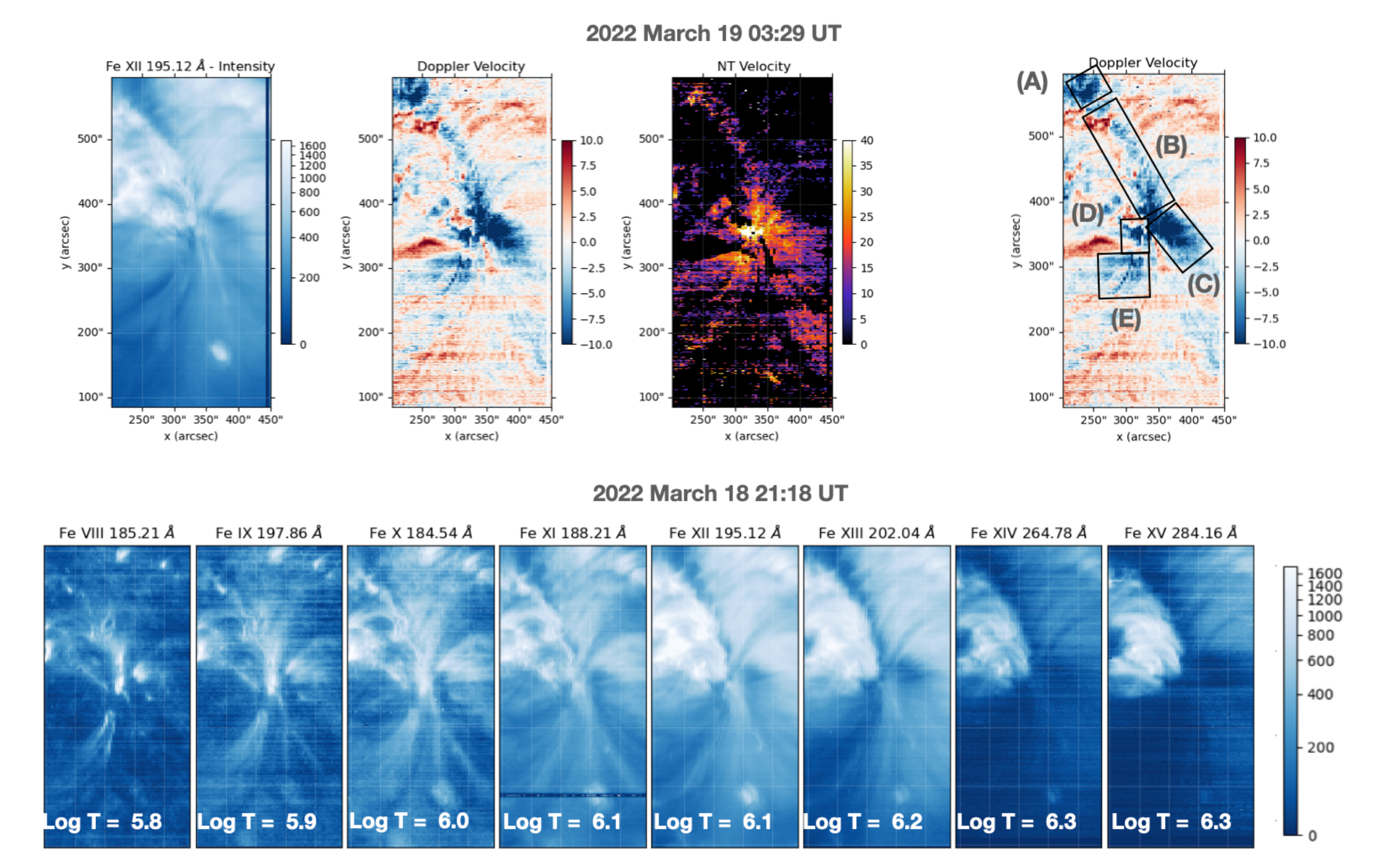}
\caption{Hinode/EIS \ion{Fe}{12} 195.12 $\angstrom$ intensity, Doppler velocity, and nonthermal velocity maps (top left to right) at 2022 March 19 03:29 UT and Doppler velocity map overplotted with boxes indicating regions of interest in the corridor (top right).  The date is close to the beginning of connectivity between the corridor and SO (see B1 in Figure \ref{fig:insitu}).  Intensity maps of Fe ions at 2022 March 18 21:18 UT (bottom) showing the expansion of the corridor width with temperature/height (see Section \ref{sec:eis}).  At this time, the corridor is close to solar central meridian where the corridor is best viewed.
\label{fig:eis}} 
\end{figure*}

The dark corridor is a striking feature in the hotter 193 $\angstrom$, 211 $\angstrom$, and 94 $\angstrom$ channels (see arrows in Figure \ref{fig:multi}) but less evident in the chromospheric 304 $\angstrom$ and upper transition region/lower corona 174 $\angstrom$ channels.
As the AR rotates across the solar disk, projection effects mean that some of the corridor is obscured.
This is especially the case in the 171 $\angstrom$ channel where there is a mixture of short, low-lying loops and long, fan loops in the vicinity of the corridor.
However, when the AR is close to central meridian on 2022 March 17, the full length of the corridor is clearly observed in the 193 $\angstrom$, 211 $\angstrom$, and 94 $\angstrom$ channels.
In these channels, the corridor significantly widens from where it starts at the positive sunspot until it blends into the northern polar CH extension.
The corridor persists and remains remarkably stable during the period of 5.5 days covered by the included movie, even after a filament eruption along the northern external polarity inversion line (PIL) at 2022 March 16 $\sim$13:00 UT.

\subsection{Hinode/EIS Observations}\label{sec:eis}
Hinode/EIS provided coordinated observations during the slow wind connection SOOP.
EIS tracked the positive polarity of AR 12967 from 2022 March 18--21.
Study ID 600/acronym DHB$\_$007$\_$v2 was employed for a total of 12 observations.
The 2$\arcsec$ slit was rastered in 4$\arcsec$ steps taking 60 s exposures at each of the 62 pointing positions.
A single observation took 62 min to build a FOV of 248$\arcsec\times512\arcsec$.
The N-S extent of the FOV covered the upflow region over the positive polarity and the dark corridor until the spacecraft pointing was shifted further south on 2022 March 20.
Co-temporal observations from SO's Spectral Imaging of the Coronal Environment \citep[SPICE; ][]{spice20} instrument do not cover the corridor, therefore, they are not included in this analysis. 

All spectroscopic data were processed with the EIS Python Analysis Code (EISPAC; \url{https://github.com/USNavalResearchLaboratory/eispac/}) software package using Python compatible level1 HDF5 files available at \url{https://eis.nrl.navy.mil}. 
The Fe ion emission lines were fitted with either single or double Gaussian functions applying the template files included in EISPAC.
Single Gaussian functions were fitted to all lines with the exception of the blended \ion{Fe}{11} 188.21 $\angstrom$, \ion{Fe}{12} 195.12 $\angstrom$, and \ion{Fe}{15} 284.16 $\angstrom$ lines.

The top panel of Figure \ref{fig:eis} shows a sample of the \ion{Fe}{12} 195.12 $\angstrom$ emission line intensity, Doppler velocity, and nonthermal velocity maps at 2022 March 19 03:29 UT when the full length of the dark corridor is evident in the EIS observations. 
This time at the Sun is roughly equivalent to when the SO spacecraft first encountered the slow SW stream taking into account the magnetic connectivity and the propagation time of the solar wind (see Section \ref{sec:insitu}).
Consistent with the observed stability of the dark corridor in images from the hotter AIA channels, the Doppler and nonthermal velocities for each of the 12 observations are very similar to those shown in the top panel of Figure \ref{fig:eis}.
Blue-shifted upflows in the range = [-20, 0] km s$^{-1}$ are observed in the Doppler velocity map over the positive polarity of AR 12967, along the dark corridor of positive field.
Nonthermal velocities within the upflow regions are in the range = [0, 44] km s$^{-1}$.

Table \ref{table:EISregions} contains measurements of the plasma and magnetic field parameters for subregions defined in the \ion{Fe}{12} 195.12 $\angstrom$ Doppler velocity map in the top right panel of Figure \ref{fig:eis}.
Only those pixels containing Doppler upflows were considered for each parameter; pixels with downflows were masked. 
Values were determined using the summed spectra within a region for each emission line to improve the signal-to-noise. 
The parameters are: FIP bias, Doppler and nonthermal velocities, plasma density using \ion{Fe}{13} ions,  temperature of the peak differential emission measure (DEM), and magnetic flux density (from the corresponding HMI magnetogram). 
The temperatures corresponding to the peak DEMs and \ion{Si}{10} 258.37 $\angstrom$ -- \ion{S}{10} 264.22 $\angstrom$ FIP bias were calculated using the method of \cite{brooks11} and \cite{brooks15}. 

Region A is located in the relatively broad northern section of the corridor.
The broader width of the region and its low magnetic flux density are more consistent with CH or polar CH extensions.
As expected, it has photospheric FIP bias with low upflow and nonthermal velocities \citep[e.g.][]{harra15,fazakerley16}. The density is slightly lower in A than in regions B and D and similar to the other two regions, however, there is little contrast in density among the five regions. 
The temperatures corresponding to the peak DEMs are similar across the regions with log $T$ in the range = [6.0, 6.2].
Where the corridor is narrow in region B, the FIP bias as well as the upflow and nonthermal velocities are comparable to regions C--E within the AR.
However, the magnetic flux there is in between CH and AR values.
Overall, the plasma parameters measured in the corridor are closer to those found at the edges of ARs \citep[e.g.][]{doschek08,harra08,brooks11,brooks16} than in CHs.

The images from different AIA channels show that the dark corridor is expanding with height/temperature in the solar atmosphere.
In order to verify whether this is the case, EIS intensity maps of emission lines sensitive to narrower temperature ranges are plotted for each ion from \ion{Fe}{8} 185.21 $\angstrom$ to \ion{Fe}{15} 284.16 $\angstrom$ at 2022 March 18 21:18 UT in the lower panel of Figure \ref{fig:eis}.
At this time, the corridor is close to the solar central meridian so that projection effects are minimized. 
The Fe emission lines represent log $T$ in the range = [5.8, 6.3] (see the values written at the bottom of the maps in the figure).
The corridor is not clearly observed in the \ion{Fe}{8} intensity map.
In the northern section of the \ion{Fe}{9} map, the dark corridor becomes evident though it is short and narrow; it continues to lengthen and widen in the intensity maps as the log $T$ increases from 5.9 to 6.3.

\begin{table}[ht!]
	\centering
	\caption{Plasma and magnetic parameters in box regions defined in the upper right of Figure \ref{fig:eis} with the \ion{Fe}{12} 195.12 $\angstrom$ Doppler velocity map at 2022 March 19 03:29 UT (see Section \ref{sec:eis}). }
	\label{tab_regions}
	\begin{tabular}{lccccc}
		\hline
		  Boxes & A  & B & C & D & E\\
		\hline
		FIP Bias & 1.2 & 2.0 & 2.1 & 1.8 & 2.1\\
		Doppler Velocity (km s$^{-1}$)& -4.1 & -11.3 & -12.2 & -13.3 & -10.1\\
        Nonthermal Velocity (km s$^{-1}$) & 19.7 & 25.6 & 28.1 & 34.3 & 25.3\\
        Log$_{10}$ Density (cm$^{-3}$)& 8.3 & 8.6 & 8.2 & 8.5 & 8.3\\
        Log$_{10}$ Temperature (K)& 6.2 & 6.2 & 6.2 & 6.0 & 6.2\\
        Peak DEM (10$^{21}$ cm$^{-5}$ K$^{-1}$)& 2.4 & 5.7 & 3.1 & 6.3 & 1.8\\
        Magnetic Flux Density B (G) &14.4&23.2&65.9&76.4&17.7\\
        
				\hline
			\end{tabular}
	\label{table:EISregions}
\end{table}

\begin{figure}[t!]
\epsscale{0.6}
\plotone{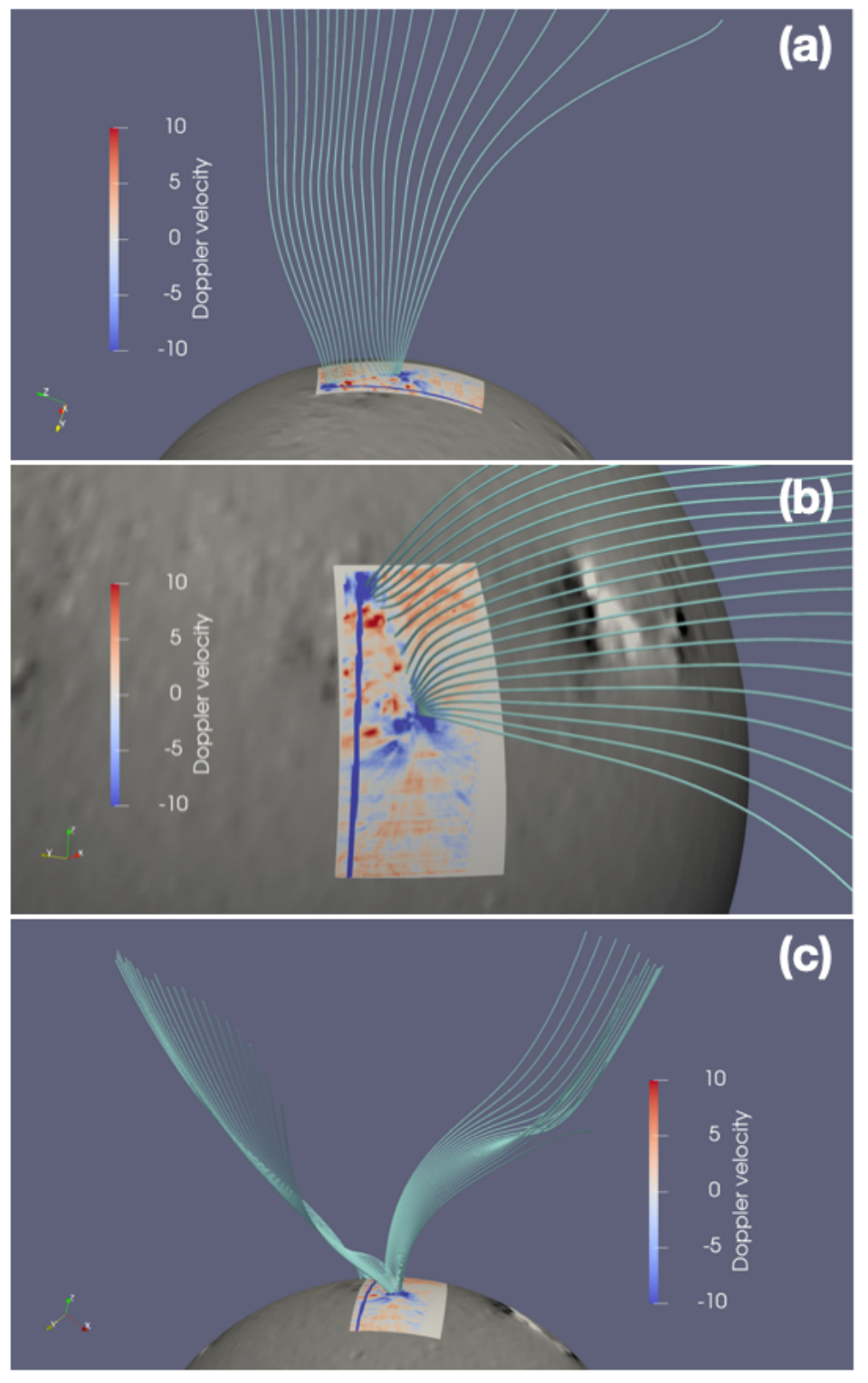}
\caption{PFSS extrapolation with Hinode/EIS \ion{Fe}{12} 195.12 $\angstrom$ Doppler velocity map at 2022 March 18 19:42 UT overlaid on SDO/HMI LOS magnetogram.  Panels (a)--(c) show open field lines from varying view points of the dark corridor (see Section \ref{sec:pfss}).
The viewpoints are: top view (a), side view (b), and along the corridor from south to north (c). In panels (a,b) field lines are set on the central part of the corridor, and on both sides of it in panel (c).
\label{fig:pfss}} 
\end{figure}

\begin{figure}[th!]
\epsscale{0.6}
\plotone{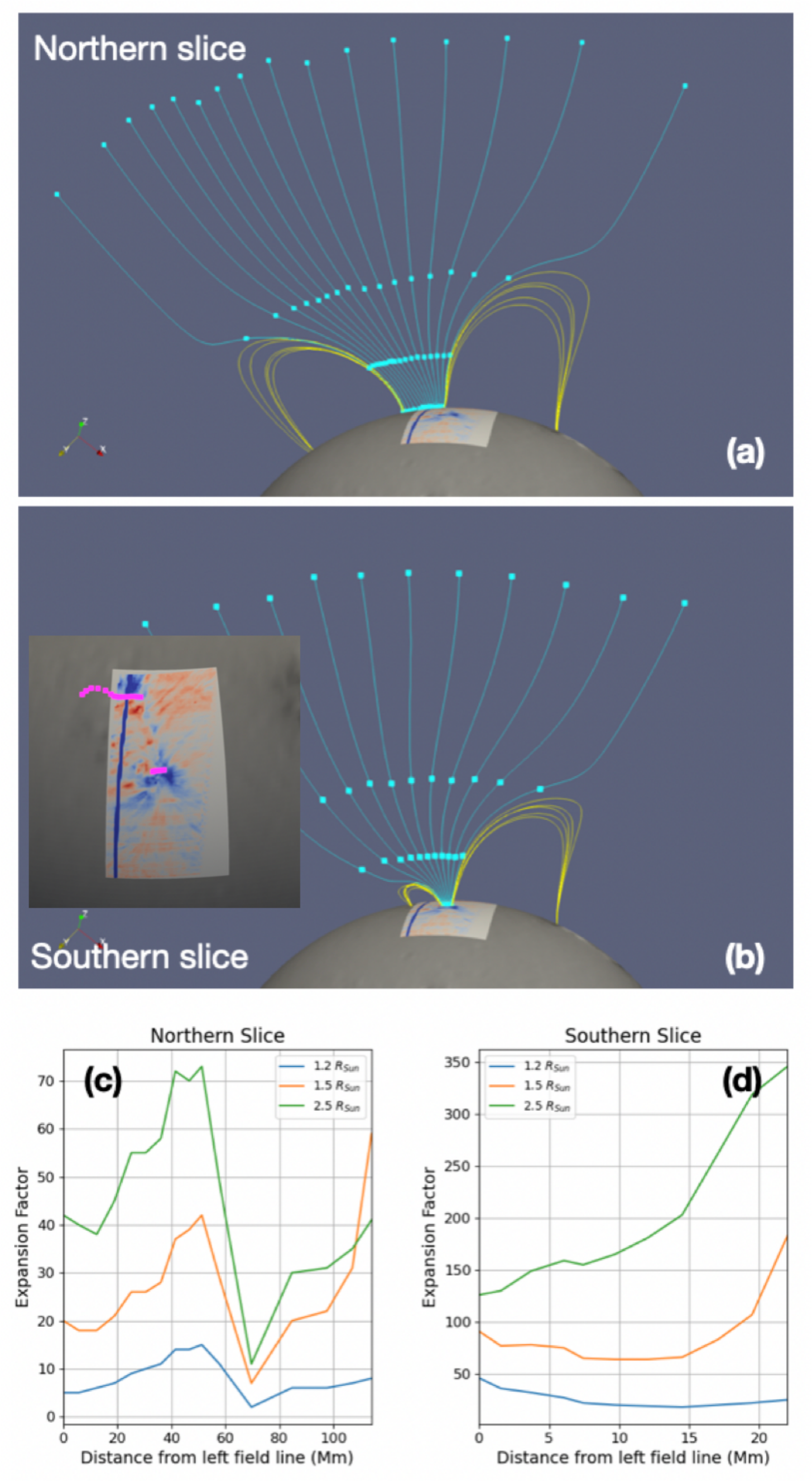}
\caption{Field lines along slices orthogonal to the corridor of upflows in the north (a) and south (b). Blue/yellow field lines represent open/closed field. Pink dots in the inset of panel (b) identify the locations of the slices. Plots of expansion factors for northern (c) and southern (d) slices vs. distance in Mm between footpoints of field lines in the photosphere starting from the left.  The three square symbols along each field line, in panels (a) and (b), correspond to radius of 1.2, 1.5, 2.5 $\Rsun$ and the expansion factor at each of the heights is plotted in blue, orange, green, respectively, in the lower panels. (see Section \ref{sec:pfss}). 
\label{fig:expfact}} 
\end{figure}

\section{Magnetic Field Modeling}  \label{sec:B}

\subsection{PFSS Extrapolation}\label{sec:pfss}

A potential field source surface (PFSS) extrapolation was used to investigate the large-scale coronal magnetic configuration of AR 12967 and the dark corridor. 
The PFSS extrapolation of the photospheric magnetic field at 2022 March 18 18:03 UT was created using the IDL SolarSoft PFSS package \citep{schrijver03}.
With this package, PFSS extrapolations are generated using the evolving surface flux transport (SFT) model based on assimilated HMI data and known methods for evolving the observed magnetic flux.
The SFT model tracks unipolar flux elements that are subject to differential rotation and meridional flows.
Flux elements will coalesce and cancel.
For each time step, the positions and flux amounts of the elements are output.
The HMI magnetogram image resolution is then binned down by a factor of four and all pixels within 60${\degree}$ of Sun-center are converted to flux elements.
A full-Sun flux map based on the HMI magnetogram inserted into the SFT model is used as the input into the PFSS code to create a 384$\times$192 full-Sun magnetogram.
Spherical decompositon is performed on the magnetogram to recontruct the magnetic field as a function of height.
The top of the PFSS model is the source surface, set at a radial distance $R_{\rm ss}$, defined here to be at 2.5 $\Rsun$.

Various viewpoints of the extrapolation are shown in Figure \ref{fig:pfss}.
Open field lines are overplotted on the Hinode/EIS \ion{Fe}{12} 195.12 $\angstrom$ Doppler velocity map embedded in a SDO/HMI LOS synoptic map of CR2255 closest in time to that of the EIS raster.
Open field lines are located along the corridor of blue-shifted upflows in the Doppler velocity map.
The top view in panel (a) includes field lines computed along the central section of the open field channel and the side view in panel (b) shows the coherence along the channel with only moderate meridional expansion in the northern and southern sections.  
A strong lateral expansion is present as shown in panel (c) where two rows of field lines are drawn on the sides of the open field corridor. 

Two cross-sections of the PFSS extrapolation orthogonal to the corridor, one in the north and the other in the south, show that the open field lines are bordered on either side by closed loops (Figure \ref{fig:expfact}a,b).
The field extends radially in the center of the corridor and then diverges away from the radial direction closer to the edges/closed loops.

The expansion properties of the open field are quantified by computing the expansion factor:
   \begin{equation}
   f(R) = \frac{\Rsun^2}{R^2} \frac{B(\Rsun)}{B(R)} \,,
   \end{equation}
where $R$ is the radial distance from the Sun center, $\Rsun$ is the solar radius,  $B(\Rsun)$ and $B(R)$ are the magnetic field strength at $\Rsun$ and $R$.  
With magnetic flux conservation, $f(R)$ provides the flux tube cross-section expansion corrected for the spherical expansion. With $R=R_{\rm ss}$, the radius at the source surface (set at $2.5 \Rsun$), $f(R_{\rm ss})$ is the expansion factor used in previous studies of solar wind sources \citep[e.g., ][]{wang90,wang09}.

The expansion factor $f(R)$ is plotted at three representative heights ($1.2, 1.5, 2.5~  \Rsun$) in Figure \ref{fig:expfact}c,d for the northern and southern cross-sections of the corridor. 
Cyan squares along each field line correspond to these heights.
The X-axis in each plot is the distance in Mm in the photosphere of each field line from the left (eastern)-most one.  
In general, the width of the open field corridor decreases and the asymmetry of the closed field increases from north to south.
The northern slice has an expansion that is smaller in the central part with respect to the lateral sides.  
At the edges, the open flux can expand rapidly when the top of the lateral closed field is reached.  This magnetic field behavior is quantified in Figure \ref{fig:expfact}c.
The plot of $f(R)$ shows a smaller expansion at low heights while reaching larger values at greater heights on both sides. 

Expansion properties change 
as the surrounding closed field and the width of the corridor vary (see abscissae at the bottom of panels (c) and (d)).   
An extreme in field line asymmetry is reached along the southern slice (Figure \ref{fig:expfact}b) where the small size of eastern arcade field allows an expansion of the corridor field at lower heights ($\lesssim 1.2 \Rsun$ in Figure \ref{fig:expfact}d).  
In contrast, the much larger western arcade field limits the corridor expansion at low heights but when the maximum height of the closed field is exceeded (approx 1.5 $\Rsun$), the open field has a bigger volume into which it can expand until it reaches pressure balance with nearby open field. This leads to larger $f(R_{\rm ss})$ on the western side than on the eastern side.  

Previous studies have shown that different expansion profiles, in particular $f(R_{\rm ss})$, have direct implications for the acceleration profiles of the solar wind, especially its terminal speed \citep{wang09,pinto16,pinto17}.  
Depending on which flux tube is crossed by SO, different solar wind properties, such as plasma speed, are to be expected with in situ measurements.

\begin{figure*}[t]
\epsscale{1.05}
\plotone{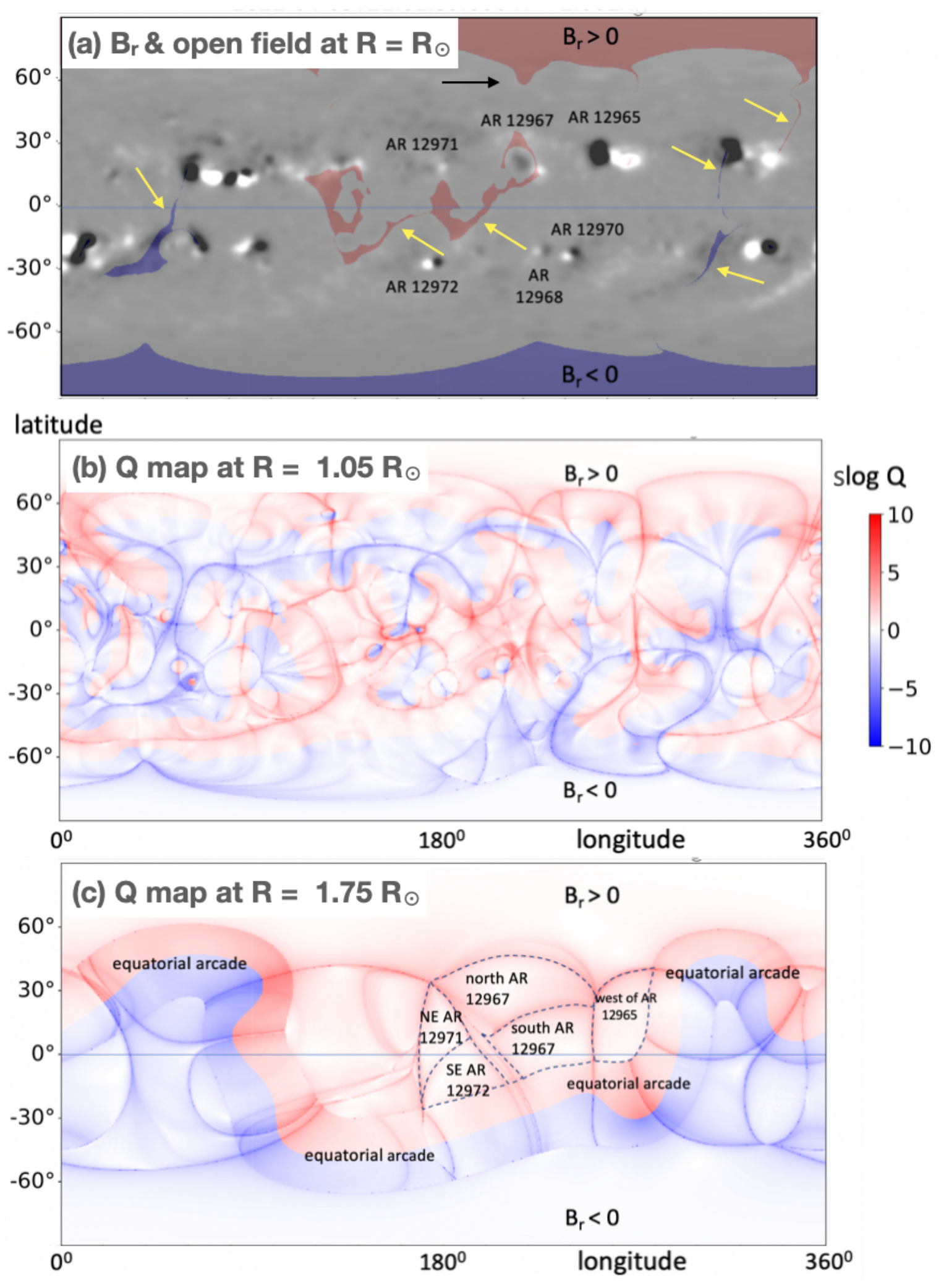}
\caption{Full synoptic maps from \url{http://hmi.stanford.edu/QMap/} (see Section \ref{sec:qmap}).
(a) Full synoptic SDO/HMI magnetogram showing photospheric footpoints of open field (red/blue is positive/negative smoothed radial B field component), 
(b) full synoptic $Q$-map at a height of 1.05 $\Rsun$ for Carrington rotation 2255, where $Q$ is the squashing factor, and 
(c) full synoptic $Q$-map at a height of 1.75 $\Rsun$ with dashed lines superposed on high $Q$ values showing the limits of open field regions in the vicinity of AR 12967.  The `equatorial' arcade is visible as a red/blue serpentine-like ribbon surrounding the Sun (beneath the heliospheric current sheet).  Black arrow in (a) indicates the northern polar CH extending towards AR 12967.  Yellow arrows in (a) identify many of the S-web corridors in the synoptic map.  
slog Q is defined by sign(B$_r$) $\times$ log [Q/2 + (Q$^{2}$/4 - 1)$^{0.5}$].
Large values of slog Q define the boundaries between regions of similar connectivities.
\label{fig:qmaps_large}} 
\end{figure*}

\begin{figure}[t]
\epsscale{0.9}
\plotone{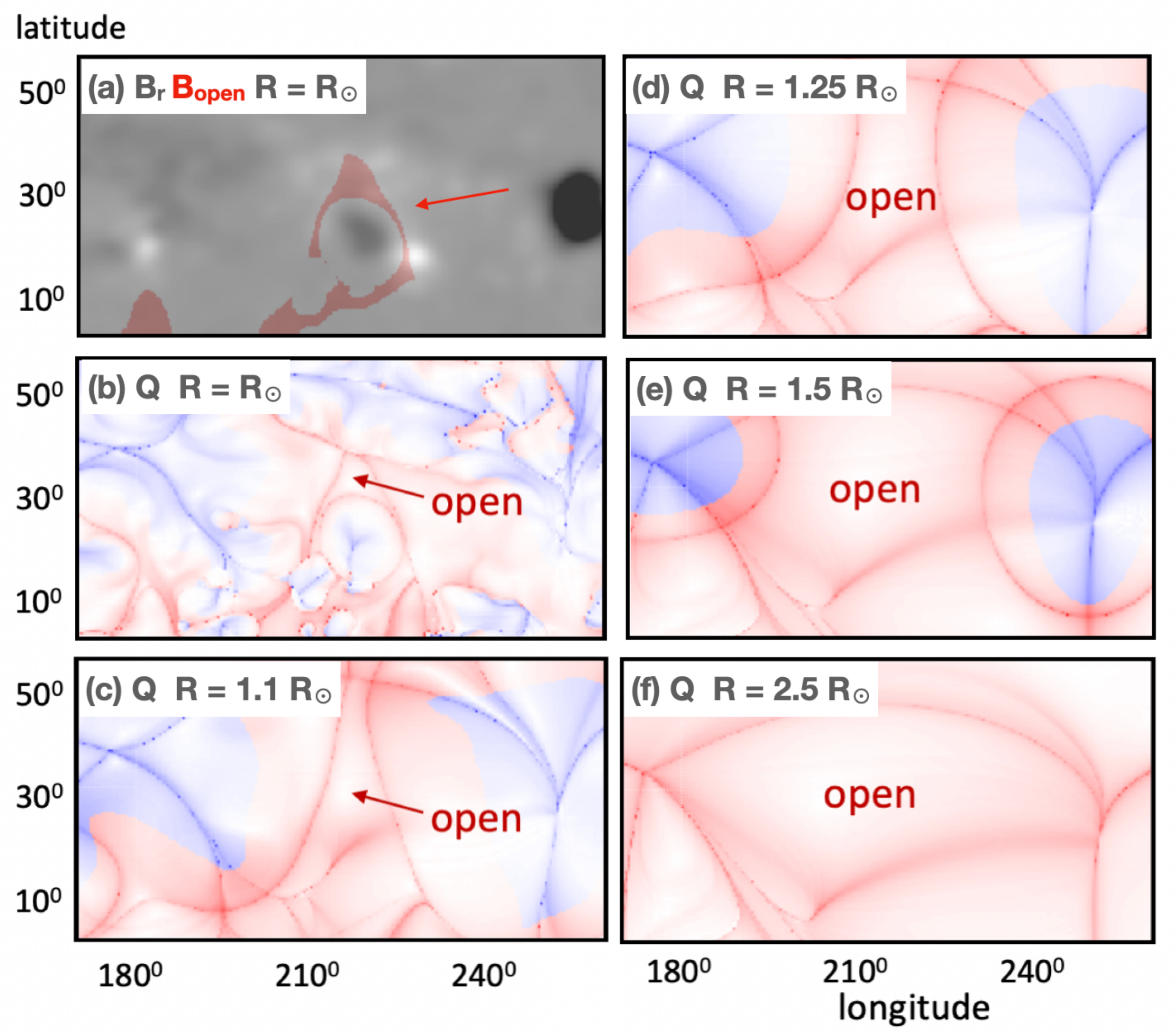}
\caption{(a) Zoomed synoptic SDO/HMI magnetogram showing photospheric footpoints of open field (red/blue is positive/negative smoothed radial B field component) around AR 12967. (b)--(f) Zoomed $Q$-maps at heights ranging from the photosphere at $\Rsun$ to the source surface at 2.5 $\Rsun$.  Dark colors in $Q$-maps represent high $Q$ values (see the scale to the right of Figure \ref{fig:qmaps_large}b).  Red arrow in (a) indicates the open field corridor. 
\label{fig:qmaps_zoomed}} 
\end{figure}

\subsection{$Q$-Maps}\label{sec:qmap} 

The open flux corridor is bordered by a drastic change in magnetic connectivity (Figure \ref{fig:expfact}) with field lines changing from reaching the source surface to reaching back to the photosphere as closed field. 
Such a magnetic configuration is typical of the presence of a separatrix and more generally of QSLs. 
Initially, QSLs were defined in closed field regions to understand the location of flare ribbons in magnetic configurations where there are no magnetic bald patches or null points present i.e. where the magnetic topology is a priori as simple as a magnetic arcade.  
In fact, for ARs, typically thin coronal volumes are present where the magnetic connectivity changes are extreme. 
These thin regions define QSLs \citep{demoulin96}. 
In the limit where the connectivity change is discontinuous, a QSL reduces to a separatrix.
A convenient way to identify strong QSLs is by using the squashing factor, $Q$, which quantifies the change in the magnetic connectivities at a boundary \citep{titov02}.  
More generally, $Q$ is a geometric measure of the degree to which neighboring field lines diverge in a 3D volume.

The concept of QSLs was extended to open/closed magnetic configurations by considering the magnetic connectivity between two surfaces, e.g. the photosphere and the source surface \citep{titov11} as was previously done for straightened  coronal configurations \citep[e.g.][]{milano99}.  
Since field lines extending outward from the source surface are supposed to follow a Parker spiral, there is no change in the $Q$ maps for a surface more distant than the source surface.

Synoptic maps of $Q$ for each Carrington rotation (CR) are available from the SDO's Joint Science Operations Center (JSOC, see \url{http://hmi.stanford.edu/QMap/}). 
Figure \ref{fig:qmaps_large}a shows the SDO/HMI synoptic magnetogram for CR 2255 with the photospheric footpoints ($r=\Rsun$) of open field in red/blue designating positive/negative radial B field components.  
The main ARs on the Sun nearby AR 12967 are labeled.
Open positive field encompasses the negative polarity of AR 12967 in the configuration of a classic pseudo-streamer \citep{wang07}.  
In the northern direction, there is an extension of the northern polar CH indicated by the black arrow in the figure.
It is plausible that the extension is connected to the open field region by a thin channel \citep{antiochos11} that is not detectable due to the low spatial resolution of the connectivity computations. 

Even though magnetic synoptic maps have highly degraded spatial resolution compared to the original HMI magnetograms and with the spatial variations further smoothed with height by a potential field extrapolation, the synoptic $Q$ map at a height of 1.05 $\Rsun$ in Figure \ref{fig:qmaps_large}b gives a sense of the complexity of the magnetic connectivities.
This is due to the fact that the photospheric magnetic field of numerous sources are linked together.
Local magnetic flux imbalance, such as with the negative polarity of AR 12967, are typically associated to closed magnetic connections to the surrounding field of the dominant opposite polarity.   
It results in a round flower-like $Q$ pattern, with the `petal' boundaries defined by high $Q$ values (see e.g. Figure \ref{fig:qmaps_large}b and Figure \ref{fig:qmaps_zoomed}d,e).
The boundaries separate the central polarity connecting to surrounding opposite polarities of larger magnetic flux. 
This is the case around the negative polarity of AR 12967, and large scale examples are also present in Figure \ref{fig:qmaps_large}b (e.g. ARs 12965 and 12970). 
Boundaries of high $Q$ are the same as the giant arcs of the S-web mentioned in the introduction.
The $Q$-maps in Figure \ref{fig:qmaps_large}b,c can be compared to Figure 7a of  \cite{antiochos11}, Figure 3 of \cite{scott18}, and Extended Data Figure 4c of \cite{chitta22}, revealing comparable patterns of $Q$.

When $Q$ maps are computed at greater heights, the flower-boundaries retract
as shown in Figure \ref{fig:qmaps_large}c (R = 1.75 $\Rsun$).
At such heights, all of the closed field regions have disappeared except for a large scale arcade which encircles the Sun (below the heliospheric current sheet).  
Without the presence of the ARs, this would correspond to an equatorial arcade with a streamer above it.  
In this active period, the arcade is warped to significant latitudes by the AR flux distribution. 
The rest of the sphere surface at 1.75 $\Rsun$ in Figure \ref{fig:qmaps_large}c is covered by open field coming from different photospheric sources, so they are separated by high $Q$ values.  
The photospheric linkage is indicated by labels around AR 12967.  
At the source surface, and farther away, only these domains of different connectivities remain.  
In summary, $Q$-maps show the extension of the various sources of the solar wind (which come into contact at the high $Q$ locations). 

Figure \ref{fig:qmaps_zoomed}a is a zoomed version of the synoptic magnetogram centered on the AR target and its surroundings. 
The open field corridor introduced in Section \ref{sec:aia} is the thin strip of positive field on the western side of the AR that broadens in its northern extension. 
The closed field region  surrounding the negative polarity of AR 12967, present in Figure \ref{fig:qmaps_zoomed}b, disappears in the maps of Figure \ref{fig:qmaps_zoomed}c, which correspond to a height difference of $0.1 \Rsun$.  
For other AR polarities, the dome of closed field can extend to much greater heights such as the eastern and western ARs (ARs 12971 and 12965) where the round QSLs are still present at $r = 1.5 \Rsun$ (right and left sides of Figure \ref{fig:qmaps_zoomed}e).
It is the retraction of the closed field with height which allows the broad lateral expansion of the open field present in between the ARs. 
Figure \ref{fig:qmaps_zoomed} provides a complementary view to the lateral expansion of the open field shown in Figure \ref{fig:expfact}.

In the $Q$ map at $r = 1.1 \Rsun$ (Figure \ref{fig:qmaps_zoomed}c), the negative polarity of AR 12967 is fully covered by open field forming a triangular-like shape.
The open field region continues to expand up to $r = 2.5 \Rsun$ (Figure \ref{fig:qmaps_zoomed}d--f).
After starting as a narrow region at $r \approx \Rsun$, the corridor broadens greatly in the longitudinal direction in contrast to only marginally increasing in the latitudinal direction.
This is in agreement with the observed broadening of the corridor with increasing temperature and height in the Hinode/EIS ion intensity maps (Figure  \ref{fig:eis}).
The significant broadening with height is related to the large expansion factors found in Figure \ref{fig:expfact}.
As a consequence, a spacecraft orbiting close to the solar equatorial plane is expected to cross a wide range of longitudes with plasma originating from the corridor even if remote observations only show a thin latitudinal extent of the source region (Figures \ref{fig:multi},\ref{fig:eis}). 
Indeed, the extent of the magnetic field at $r = 2.5 \Rsun$ is up to about $60^{\circ}$ of longitude (Figure \ref{fig:qmaps_zoomed}f), which is comparable to the typical distance between neighboring ARs in Figure \ref{fig:qmaps_large}a.

\begin{figure*}[t]
\epsscale{1.2}
\plotone{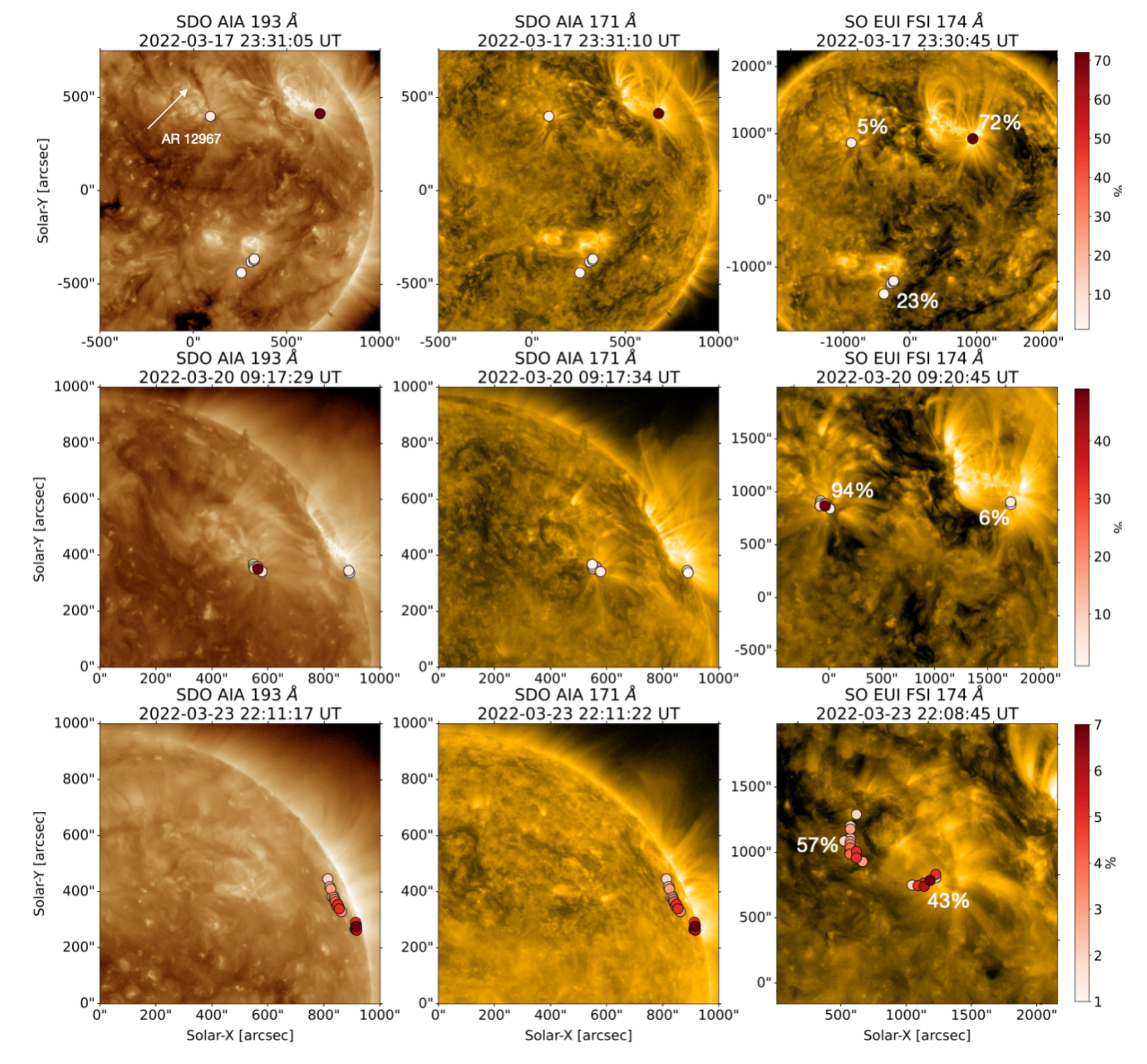}
\caption{SDO/AIA 193 $\angstrom$, SDO/AIA 171 $\angstrom$, SO/EUI FSI 174 $\angstrom$ images (left to right) on 2022 March 17, 20, 23 (top to bottom, Sun times) showing connectivities from the MCT at Sun times. This is between 1.5 and 2.25 days earlier than spacecraft times for this dataset.  Separation angle between SDO and SO is 40\degree . 
Colors of dots show the probability of the connectivity (see color coding on the right side and Section \ref{sec:connect}).  The percentages are the probabilities for each region.  White arrow indicates the open field corridor of S-web.
\label{fig:connect_eui}} 
\end{figure*}

\begin{figure*}[th!]
\epsscale{1.1}
\plotone{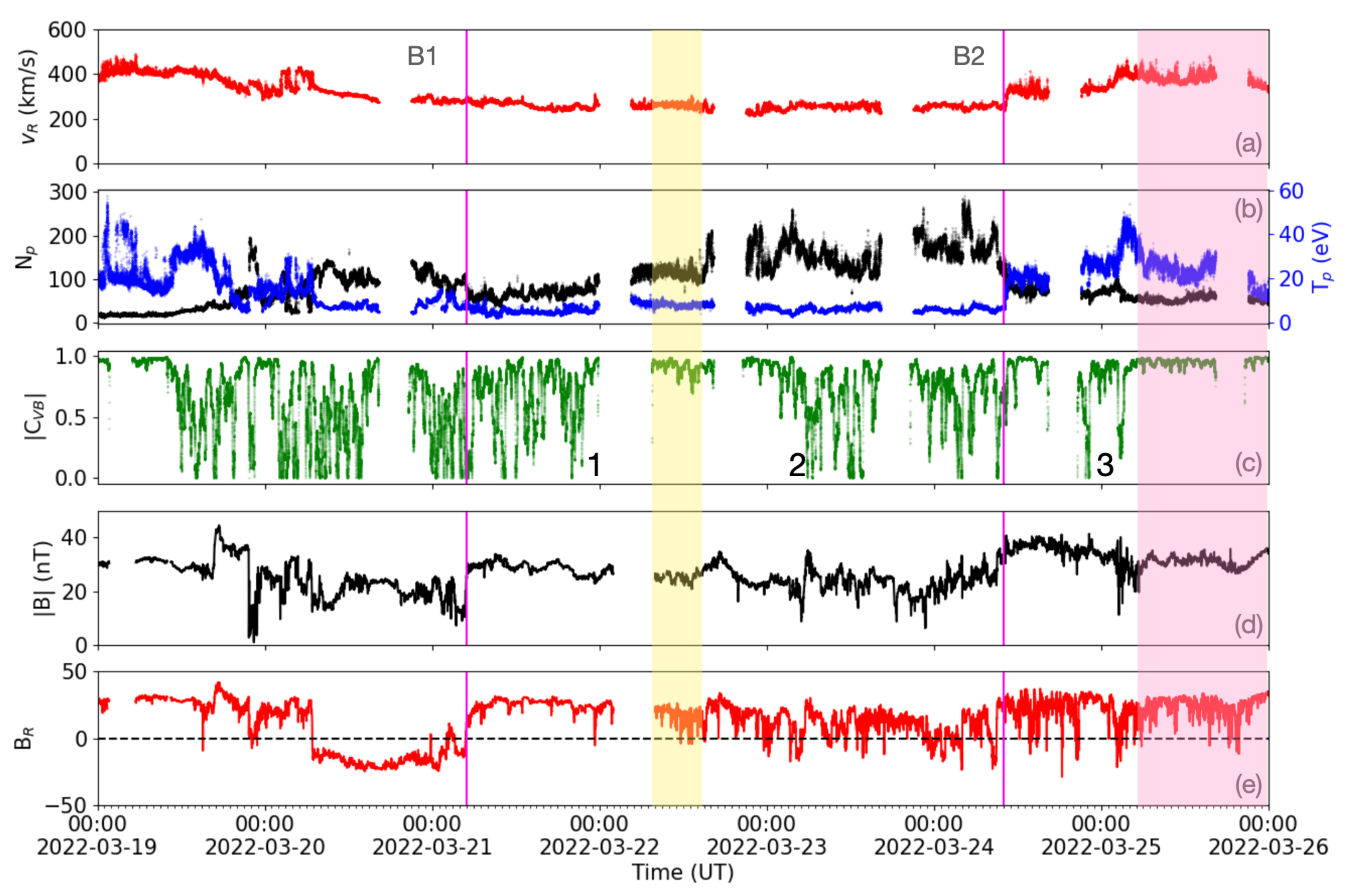}
\caption{Solar Orbiter SWA-PAS (panels (a)--(b)) and MAG (panels (d)--(e)) in situ data:  (a) proton velocity $V_{R}$ (km s$^{-1}$) in red, (b) proton number density N$_{p}$ (cm$^{-3}$) in black and proton temperature $T$ (eV) in blue, (c) absolute value of correlation coefficient |C$_{vb}$| in green, (d) total magnetic field |$B$| (nT), and (e) radial magnetic field component $B_R$ (nT) in red. |C$_{vb}$| is computed over a running window of 30 min.  The pink vertical lines B1 and B2 denote the start and end of the interval of connectivity with the narrow corridor. 
Shaded areas indicate periods of moderate Alfv\'enicity.  Times are at the spacecraft which is between 1.5 and 2.25 days later than at Sun times (cf. Figure \ref{fig:connect_eui}).
Numbers correspond to locations in the extrapolation in Figure \ref{fig:mid_slice}. These locations indicate the inverted U-shaped Alfv\'enicity pattern discussed in the text. }
\label{fig:insitu} 
\end{figure*}

\section{ Magnetic Connectivity}  \label{sec:connect}

The Magnetic Connectivity Tool (MCT) \citep{rouillard20} was used to connect the in situ measurements at SO to the possible on-disk source regions. 
The MCT models the corona with a potential field computed from a synoptic map of the photospheric field, then it performs a PFSS reconstruction similar to what is described in Section \ref{sec:pfss}.   
The photospheric field could be either an NSO/GONG synoptic map derived directly from observed line of sight magnetograms or an ADAPT synoptic map using flux transport schemes to evolve the observed magnetic field near the central meridian to the required time \citep{hickmann15}. Examples of NSO and ADAPT synoptic magnetograms are provided in the Appendix.  
For the PFSS reconstruction, the magnetic field is computed up to the source surface ($R=R_{\rm ss}$) where the magnetic field is supposed to be radial.  
At larger distances, the interplanetary magnetic field is assumed to be a Parker spiral with a constant solar wind speed. 

Based on ADAPT synoptic maps, the MCT establishes partial or full connectivity between the positive polarity region of AR 12967 and SO during the period 2022 March 20--26 (spacecraft time corresponding to 2022 March 18--24 at Sun time i.e. a solar wind parcel departs from the Sun at the Sun time and arrives at  SO at the spacecraft time). 
From 2022 March 20--22, connectivity is found to be to the positive fields of both AR 12967 and AR 12965 (to the west).
A full and stable connection to AR 12967 is achieved only after the beginning of 2022 March 22 (spacecraft time).
Partial disconnection begins early on 2022 March 25 when SO is connected to the positive field on either side of the negative polarity of AR 12967.
SO is fully disconnected from AR 12967 from 2022 March 27.

Figure \ref{fig:connect_eui} illustrates the connectivities and their probabilities from MCT ADAPT maps during each interval (2022 March 17, 20, 23 at Sun times) plotted on SDO/AIA 193 $\angstrom$, SDO/AIA 171 $\angstrom$, and SO EUI/FSI 174 $\angstrom$ 
maps.  
Distributions of connectivity points are calculated using multiple field lines generated by sampling an uncertainty ellipse region around the target spacecraft position and then remapping the respective probability densities across the interplanetary and coronal magnetic field \citetext{Rui Pinto, 2022, private communication}.
The percentages in the SO/EUI column of images are the probabilities for each grouping of dots.
It should be noted that connectivity depends on a number of factors including: the data input (e.g. types of magnetograms, cadence of magnetograms) at the base of the PFSS reconstruction, the height of $R_{\rm ss}$ \citep{panasenco20,koukras23}, the wind speed and its acceleration profile \citep[e.g. as deduced from a statistical analysis,][]{Dakeyo22}, and the inclusion of more realistic physics. 
Differences in the timing and connectivities produced by varying the input parameters demonstrate the limitations of setting the connectivity with such basic models (PFSS and Parker spiral).  
Nonetheless, this method provides clear hints of the magnetic connectivity between AR 12967 and SO, while the timing is refined below with the in situ data. 

\section{In situ Observations}  \label{sec:insitu}

SW data were obtained from SWA-PAS and MAG in situ instruments on board SO.
Figure \ref{fig:insitu} shows SWA-PAS 4 sec time resolution radial velocity V$_{R}$, proton number density N$_{p}$ (black) and proton temperature T$_{p}$ (blue) data in panels (a)--(b) and total $|B|$ and radial B$_{R}$ MAG magnetic field data in panels (d)--(e) for the period 2022 March 19--25. 
As is also the case for the synoptic maps, $B_R$ in the outward/inward direction from the Sun is defined as positive/negative field.
Panel (c) is the absolute value of the correlation coefficient C$_{vb}$ between $V_N$ and $b_N$ (with $b_N$ being the normal component of the magnetic field in Alfv\'en units, $b_N = B_N/(4 \pi \rho)^{1/2}$, with $\rho$ being the mass density and $B_N$ the normal component of the magnetic field) computed over a running window of 30 min, which is a typical Alfv\'enic scale in the solar wind \citep{tu95,bavassano98}.
All in situ data are plotted against measurement time at the SO spacecraft (in the range = [0.32, 0.38] au).  

It should be noted that for solar wind velocities below $\sim$ 275--300 km s$^{-1}$, a part of the velocity distribution function may fall into the low energy measurement range of the SWA-PAS sensor.  It is known that there is an instrumental issue at these low energies at which the counting efficiency is reduced.  This means that under these unusual circumstances the low energy ions can be undersampled.  If a significant part of the distribution lies in this low energy range, then there is likely to be an effect on the subsequent calculation of the moments (velocity, density, and temperature) of the distribution.  A quality factor parameter is  provided by the sensor teams as an assessment on the potential level of impacted data points:
A quality factor of <0.2 means that there are no significant issues with the derived moments; <2.0 means that there can be some under-sampling of the low-energy data; >2.0 means that under-sampling may be significant and caution must be used in quantitative interpretation of the data \citetext{A. Fedorov, 2023, private communication}\footnote{See also 'SWA/PAS Tutorial' at \url{https://drive.google.com/drive/folders/1d2Y-G0BiAyAyQqTXL6x9zI39PosoWay\_}}.  
The data in interval B1--B2 (Figure \ref{fig:insitu}) has a median quality factor = 1.6 and $\sigma$ = 1.6.  
The high densities of this very slow solar wind period may be underestimated. 

Magnetic mapping techniques produce a relatively wide window of partial/full connectivity between the open field corridor of AR 12967 and SO (Section \ref{sec:connect}).
The sharp transitions observed in the in situ data provide for better defined boundaries of this interval (B1--B2 in Figure \ref{fig:insitu}).
$B_R$ (panel (e)) changes from negative to positive at 2022 March 21 $\sim$04:00 UT. 
It remains clearly positive on the 22$^{nd}$ and generally positive on the 23$^{rd}$ though showing far more variability.
However, early on 2022 March 24, $B_R$ has low mean positive values with several excursions into the negative domain before becoming suddenly mostly positive after B2.
$|B|$ slightly increases at this transition and thereafter has a nearly constant value.  
Abrupt changes in both $B_R$ and $|B|$ are the result of a change in magnetic connectivities.  
Boundary B2 is further confirmed by the fact that the plasma parameters show discontinuities at the time associated with this boundary in Figure \ref{fig:insitu}a,b.

The time period covered between B1 and B2 is from 2022 March 21 $\sim$04:00 UT to 2022 March 24 $\sim$09:00 UT, so with a duration of 3 days and 5 hours, slightly less than the duration of about 3.5 days of stable and unique connection to the western positive polarity of AR 12967 given by the mapping from both ADAPT and NSO synoptic maps.
During the B1--B2 interval, upflows along the corridor observed on the Sun (Figure \ref{fig:eis}) are expected to be outflows as determined from the magnetic field extrapolations (Section \ref{sec:pfss}) and in situ data. 

The highly variable density of the slow SW stream in the B1--B2 interval increases from approximately [60--80] cm$^{-3}$ at B1 to [150--200] cm$^{-3}$ at B2, with spikes approaching 300 cm$^{-3}$.
The level of $N_{p}$ is high since 100 cm$^{-3}$ at a distance of 0.35 au translates to 12 cm$^{-3}$ at 1 au, assuming a classical $r^{-2}$ dependence, therefore this stream would have a density between 7 and 25 cm$^{-3}$ at 1 au.  
$N_{p}$ is comparable to the denser slow SW measured at 0.35 au by Helios \citep[see Figure 5 of][]{maksimovic20}. Furthermore, apart from local small spikes, $V_{R}$ is remarkably uniform. 
It is close to the slowest wind velocities measured at this solar distance with $V_R$ around 300 km s$^{-1}$.  
The proton temperature $T_{p}$ of a few eV ($\sim$$10^4$K) is similar to the Helios slow SW measurements at about 0.3 au \citep{perrone20}.
This is consistent with the known correlation between $T$ and $V_R$ at 1 au \citep{lopez86,elliott05,elliot12,perrone19}.  

The magnetic field strength in the interval B1--B2 is homogeneous apart from fast spikes of lower values (Figure \ref{fig:insitu}d).  The radial component $B_R$ is mostly anti-sunward directed as expected from the positive polarity of the corridor. 
The large variability and occasional reversal events are likely to be switchbacks \citep[e.g.][]{neugebauer13,horbury18,horbury20}. 
More globally, $B_R$ is progressively decreasing to smaller values before increasing at the end of the interval close to B2. 
\begin{figure*}[t]
\epsscale{1.1}
\plotone{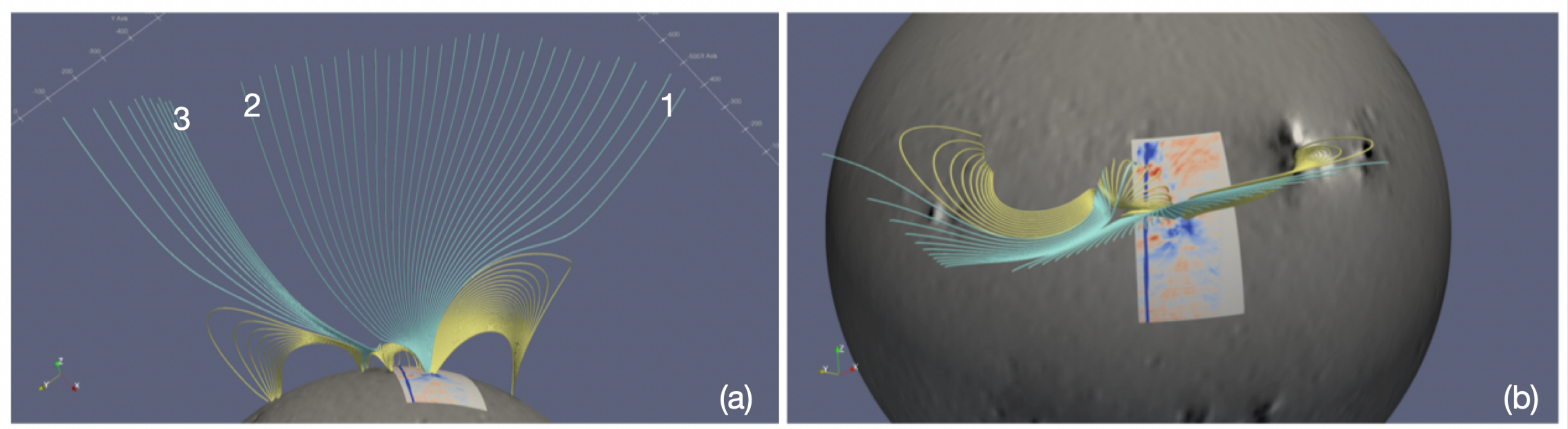}
\caption{Same PFSS extrapolation and Hinode/EIS  map as in Figure \ref{fig:pfss} from side (a) and top (b) viewpoints.  
Blue/yellow field lines represent open/closed field.
A topological dome encompasses the negative polarity of AR 12967 in (b).
Numbers correspond to those in Figure \ref{fig:insitu}.
\label{fig:mid_slice}} 
\end{figure*}

An indication of Alfv\'enicity is a strong correlation between $V$ and $B$ components, however, for the sake of simplicity, only the normal component is considered here since it is more Alfv\'enic than the radial and tangential components \citep{tu89}.
In the B1--B2 interval, |C$_{vb}$| oscillates in the range of [0, 1], with spikes below $\approx 0.5$ indicating poor Alfv\'enic correlation. 
However, on 2022 March 22 the $V$-$B$ (component) correlation is more stable and in the range = [0.7, 1.0] for approximately 1/3 of a day (see the yellow shaded area in Figure \ref{fig:insitu}c).
In addition, N$_{p}$ and $B$ are roughly constant during the Alfv\'enic period suggesting low compressibility. 
N$_{p}$ then exceeds 200 cm$^{-3}$ accompanied by rising |$B$| on either side of the second data gap indicating more mixed compressibility.
In combination, the in situ observations indicate the presence of Alfv\'enic fluctuations in the shaded region during the B1--B2 interval.
Next, from early on 2022 March 23, |C$_{vb}$| oscillates between 0 and 1 again until approximately 2022 March 25 05:00 UT, after which the slow wind stream is more Alfv\'enic (red shaded area).

After B2, SO is mainly connected to the diffuse positive polarity on the east side of AR 12967's negative polarity according to ADAPT synoptic maps. 
The open magnetic field there does not expand to the same extent as on the western side (see Figure \ref{fig:mid_slice}, which shows the magnetic field along an east-west slice of the extrapolation through the negative polarity of AR 12967).
This is indicative of a much lower expansion factor compared to the western side, in agreement with higher in situ velocities and proton temperatures and lower plasma density.
A detailed analysis of the Alfv\'enic SW intervals during the 30-day perihelion passage of SO is available in \cite{damicis23}.

Linking SW streams to their source regions is not straight forward.
In this case, the SO MCT connectivity predictions after the B1--B2 interval are widely different, depending on whether the input magnetograms are NSO or ADAPT (see images and movies in the Appendix). 
The NSO LOS magnetogram shows that the connectivity shifts from the positive polarity of AR 12967 along the channel of positive field leading to dispersed field in the southern hemisphere near AR 12972 (see Figure \ref{fig:qmaps_large}a) whereas the ADAPT connectivity prediction is to the positive field immediately to the east of embedded negative polarity of AR 12967.

The Alfv\'enicity level provided by $|C_{vb}|$ has several inverted U-shape patterns. The first one is observed from March 22 00:00 UT until March 23 05:00 UT (corresponding to 1 and 2 labels in Figure \ref{fig:insitu}), and a second one (half of an inverted U-shape) is present from March 24 03:00 UT until after March 26 00:00 UT (corresponding to 3 label in Figure \ref{fig:insitu}). 
These inverted U-shape patterns are consistent with SO changing connectivity from the CH-like field on the western side of the AR 12967 negative polarity to the CH-like field on the eastern side.  
More precisely, the locations of these numbers are shown in the coronal extrapolation of Figure \ref{fig:mid_slice}.
It matches the ADAPT connectivity predication rather than that of the NSO (where the connection is extending progressively to the channel of positive field to the southeast of AR 12967).  
The central part of the corridors has a lower expansion factor (e.g. Figure \ref{fig:expfact}) and is associated with $|C_{vb}|$ close to 1.  
The sharp fall in Alfv\'enicity at 1--2 and 3 in Figure \ref{fig:insitu} is comparable with what is observed in equatorial CHs \citep{wang19,damicis21a}.
These inverted U-shape patterns are an important marker of the magnetic field differential expansion across open field regions, i.e. of individual open fields in the low corona where they are spatially separated.

In contrast, the NSO connectivity prediction is in between AR 12967 and AR 12972 (see Appendix and associated movies). 
The photospheric magnetic field along this path is highly dispersed compared to that of the narrow corridor, therefore, a lower flux tube expansion with height is to be expected. 
This is confirmed in the PFSS extrapolation by computing an ensemble of field lines (not shown). 
Furthermore, there is flux tube contraction with height moving away from AR 12967 toward AR 12972. 
This behavior is different than that within the corridor and the associated yellow shaded region, while the yellow and red shaded regions have broadly similar Alfv\'enicity. 
Finally, the distance between the two ARs relative to the distance across the negative polarity of AR 12967 would require a much larger time period between the yellow and red shaded regions in Figure \ref{fig:insitu}.  
We conclude that the insitu observations do not agree with the PFSS computed from NSO synoptic magnetograms.

Plasma composition and charge state are well known in situ parameters that are useful for refining/confirming connectivity, however, these data were not available from SO's perihelion.
Instead, a combination of the expansion factors in different areas of the PFSS extrapolation and Alfv\'enicity of the slow SW provide a high confidence level that the connectivity shifts from the west to east side of the negative polarity of AR 12967 as predicted by the MCT using ADAPT magnetograms.


\section{Discussion and Conclusions}
\subsection{Characterization of Narrow Corridor and its Associated Slow Wind Stream}
SO tracked the decaying AR 12967 during its solar disk crossing and obtained both remote sensing and in situ observations. 
SO was also magnetically connected to AR 12967 and its vicinity for about 5 days.
A distinguishing feature of the AR was a dark and narrow corridor observed in the EUV that expanded with height and temperature in the corona (Figures \ref{fig:multi},\ref{fig:eis}).
Plasma upflows of approximately 10 km s$^{-1}$ ran along the full extent of the corridor as observed by  Hinode/EIS.
AR 12967’s magnetic field structure has a classical pseudo-streamer configuration with the following negative polarity surrounded by positive field (its leading polarity counterpart in the southwest and remnant fields elsewhere, Figures \ref{fig:qmaps_large}-\ref{fig:qmaps_zoomed}).
The open positive field in the west is spatially coincident with the plasma upflows of the narrow corridor, implying that the upflows are outflows (Figures \ref{fig:pfss}-\ref{fig:expfact}, \ref{fig:mid_slice}).

In situ measurements at SO show that the SW stream emanating from the narrow corridor had exceptionally slow radial velocities of $\sim$300 km s$^{-1}$ with correspondingly low proton temperatures of about 5 eV and increasingly high proton density of from $\sim$ 60 to $300$ cm$^{-3}$ in the time interval defined by boundaries B1 to B2 (Figure \ref{fig:insitu}).
For approximately 1/3 day during this interval, the slow SW exhibited moderate Alfv\'enicity with the correlation coefficient between $V$ and $B$, |C$_{vb}$|, in the range [0.7, 1].
This period was preceded and followed by more variable, and non-Alfv\'enic periods, especially closer to the B1 and B2 boundaries.

Alfv\'enic slow SW was first reported by \cite{marsch81} using Helios 2 measurements at 0.3 au.
Subsequent in situ observations confirmed the presence of Alfv\'enic slow SW streams at $\sim$0.3 au \citep[e.g.][]{damicis15,bale19,stansby19,stansby20,perrone20} and at 1 au \citep[e.g.][]{damicis11,wang19,damicis19,damicis21a}.
Alfv\'enicity is a key discriminator for classifying the slow SW and identifying its source regions especially at close distances like that of SO at its perihelion \citep{stansby19,damicis21a}.
In this case, it was especially important in the absence of in situ composition data.

While proton temperatures and magnetic field flux densities are consistent with the average values measured by Helios within non-Alfv\'enic slow SW at a distance of 0.3--0.35 au, the velocity and density parameters detected at SO are extreme \citep[e.g. see Table 1 of ][]{perrone20}.
In fact, they are comparable to the values of the so-called very slow SW described by \cite{sanchez16} where velocities <300 km s$^{-1}$ are associated with much higher and more variable densities in contrast with those reported by \cite{perrone20}.
The very slow SW was observed by Helios 1 and 2 at distances mainly $<0.5$ au and not above $0.7$ au.
\cite{sanchez16} proposed that the origin of the very slow SW may be small and isolated CHs, i.e. equatorial CHs, that become more widespread during active periods of the solar cycle.

Exceedingly high densities combined with very low velocities provide crucial information for further refining the source region--spacecraft connectivity during the B1--B2 interval.
SW velocities of $\sim$300 km s$^{-1}$ are more likely to originate at the southern end of the narrow corridor where $f(R_{\rm ss})$ is an order of magnitude higher compared to that of the north (see Figure \ref{fig:expfact}c,d). 
Furthermore, the southern end of the corridor is adjacent to the core of AR 12967.
Typically, the core loops connecting opposite polarities  have the highest density \citep[e.g.][]{tripathi08} and coronal FIP bias \citep[e.g.][]{gdz14,baker15,baker21} values within the various loop populations of ARs.
The in situ plasma during B1--B2 is more characteristic of the core loops in the south compared to the CH and quiet Sun found at the northern end of the corridor.
Finally, the positive polarity of AR 12967 is a natural location for  reconnection to take place between core loops and nearby open field \citep{baker09} so that the plasma can be transferred from the closed core loops to nearby open field.

A comparison of the plasma parameters and magnetic field within the narrow corridor and its associated SW stream with that of equatorial CHs provides insight as to where they are located on the scale of CHs.
\cite{wang19} compiled the basic properties of 14 equatorial CHs and their associated slow wind streams ($\sim$350--450 km s$^{-1}$) at 1 au from 2014--2017 (during solar maximum/beginning of decay phase of cycle 24).
They found the following:  1.  longitudinal widths $\sim$ [3$^{\circ}$, 10$^{\circ}$]; 2. relatively high Alfv\'enicity |C$_{vb}$| $\sim$[0.6, 0.7] that tended to fall sharply approaching the edges or boundaries of the slow SW streams; 3.  footpoint magnetic field strengths $\sim$[15, 30] G; 4.  proton temperatures $\sim$[1, 3] eV; 5.  large expansion factors >9.
These results, where applicable, correspond to those of \cite{damicis19} who identified equatorial CHs as the source regions of highly Alfv\'enic slow SW at 1 au.

The comparable values for key properties of the narrow corridor and its associated very slow SW are more extreme than those of equatorial CHs:
the longitudinal width of the corridor in the 1.5 MK corona is up to 2 orders of magnitude smaller, the expansion factors are as much as an order of magnitude larger in the south, and the SW velocities are lower by $\sim$100 km s$^{-1}$.
Alfv\'enicity in the B1--B2 interval possibly reaches higher levels compared to the equatorial CHs in the study of \cite{wang19}, however, the extent of Alfv\'enicity appears to be shorter.
The extremely high expansion factor related to this short period of Alfv\'enicity is expected as there is low expansion only in the center of the flux tube compared to the edges. 
Closer to the edges, the field becomes more curved so that more of the Alfv\'en waves are reflected back rather than traveling outward along the open field. 

\subsection{Narrow Corridor and the S-web}
The narrow corridor observed by Hinode/EIS  has key features predicted by the S-web model.
First, at typical coronal temperatures, it is a few orders of magnitude thinner in the longitudinal direction compared to typical equatorial CHs that were confirmed source regions of the slow SW.
Second, the corridor is conjectured to be topologically robust \citep{antiochos07,antiochos11}. 
In this case, the dark corridor was observed to be highly stable for at least seven days.
Third, the corridor expands to a latitudinal width of 60\degree\ 
at the source surface (=2.5~$\Rsun$). 
Fourth, the overall magnetic configuration is a pseudo-streamer formed of the embedded negative polarity of AR 12967
with the corridor formed within the surrounding positive field on the west.
Open field corridors are intrinsically associated with pseudo-streamers 
\citep{antiochos11,crooker12,scott19}.

Some of the extreme properties of the slow SW stream observed by SO are a direct consequence of the very nature of the S-web comprising open field corridor source regions.  
The very slow SW recounted by \cite{sanchez16} is plausibly produced by such regions.
Extremely narrow corridors are characterized by super-radial expansion and therefore very slow wind velocities.
Following on from this, Alfv\'enic slow SW is associated with over-expanded flux tubes \citep[see][and references therein]{damicis21a}.
|C$_{vb}$| was in the range [0.7, 1] for 1/3 days during the central part of B1--B2 interval.
Large variability in $B_{R}$ and occasional reversal events, i.e. switchbacks, occurred while the very slow SW was moderately Alfv\'enic.
Recent PSP and SO observations have established that switchbacks are linked to the Alfv\'enic slow wind from equatorial CHs \citep[e.g.][]{bale19,damicis21a,fedorov21}.
It is reasonable that the induced interchange reconnection events along the S-web arcs \citep{higginson17b} are a possible source of the switchbacks everpresent in the inner heliosphere \citep[][and references therein]{damicis21a} and observed in the Alfv\'enic very slow SW emanating from the narrow open field corridor.

The corridor associated with AR 12967 is not rare.
On the contrary, the full synoptic HMI magnetogram shown in Figure \ref{fig:qmaps_large}a contains a number of possible S-web corridors identified by the yellow arrows.  
There is a spectrum of widths, mainly as a consequence of the  B$_{R}$ distribution in the photosphere.
The narrowest of corridors will be beyond the spatial resolution of the magnetic models used to determine the open field regions.
Furthermore, limitations of EUV observations e.g. instruments with limited FOV, and projection effects (corridors are masked by nearby brighter regions) also contribute to the mistaken impression that they are not common features.

In conclusion, the combination of observations from SO and Hinode/EIS has made it possible to characterize the S-web narrow open field corridor source region and its slow solar wind stream.
The intrinsic topology of the corridors means that super-radial expansion of the open field is likely to yield extremely slow SW velocities, with moderate Alfv\'enic content (along the central section of the corridor), and switchbacks.
Other SW properties such as composition, charge state, and density are more likely to be governed by the surroundings of the corridors as interchange reconnection at the corridor boundaries opens up pathways for closed-field plasma of nearby quiet Sun and active regions to reach the heliosphere.

\section{Future Work}
The present study is in line with previous studies thereby improving our understanding of the link between solar and in situ observations as summarized in Section \ref{sect_Introduction}.  
Connectivity computations need to be pursued in a variety of other cases to test the statistical validity of the above conclusions.   
This can be achieved with the data presently accumulated by SO, PSP, SDO, Hinode and ground based observatories. 
For example, a natural extension of this work can be achieved during the 30-day perihelion passage of SO based on the study of \citet{damicis23}.  
PFSS extrapolation and magnetic field topology computations, as well as the outward tracing with a Parker spiral, adapted to the in situ measured velocity, are well developed tools and sufficiently light in computational time to analyze a large amount of data.   
This method has known limitations which will need to be carefully tested in a large set of cases. 
Such tests that are worth conducting include how to further constrain where to set the source surface radius \citep{panasenco20,koukras23} and how to better understand the open flux deficit \citep{linker21}.  
These can best be realised when the HCS is significantly warped so that the spacecraft successively scans different and identifiable source regions (e.g. of opposite magnetic polarity and/or of different spatial extents thereby containing different solar wind types).  
Furthermore, the transit time from the corona to the spacecraft needs to be improved. 
One possibility is to use the statistical results of the wind profiles obtained by combining Helios and PSP observations  \citep{Dakeyo22}.

The precision of the deduced connectivities with various models and photospheric observations needs to be  quantitatively evaluated in order to derive what is the best mapping function of the studied coronal configuration \citep{Badman22}.  
However, the approach used in this study as well as most other ones involve a static approach to connectivity while the corona and the solar wind are known to be dynamic over a broad range of times (seconds to years), therefore, the above analysis is only providing the steady part of the connectivities, on a time scale of a few weeks. Further improvements call for the development of MHD numerical codes (coronal and interplanetary) to deal with the temporal evolution, e.g. the formation of stream interaction regions.  
This is a long term effort which requires both an improvement in the boundary conditions (constrained by observations) and in the input physics \citep[such as the required extra heating and solar wind acceleration, e.g. see ][for more information]{perri23}. 
The already impressive harvest of SO and PSP observations, together with the development of numerical tools, will allow us to deeply improve our knowledge of coronal and interplanetary physics, which are presently summarized in \citet{Rouillard21,Raouafi23}.

\appendix
The Solar Orbiter Magnetic Connectivity Tool (MCT) described in Section \ref{sec:insitu} identifies the possible source regions of the solar wind streams. 
Sample MCT ADAPT (Figure \ref{fig:append1}) and NSO (Figure \ref{fig:append2})  input magnetograms are shown below.
The images are from included movies called MCT$\_$ADAPT.mp4 and MCT$\_$NSO.mp4.
A comparison of the two movies shows differences between the potential magnetic field and the connectivities generated from different input magnetograms.
The most striking feature is the dissimilarity in the heliospheric current sheets (HCS).  
The HCS in the ADAPT map runs from NW to SE, crossing the solar equator, whereas the one in the NSO map `wraps' around the northern (ARs 12967 and 12971) and southern (ARs 12668, 12670, and 12672) activity belts.
Until early on 2022 March 24, SO connects to both AR 12965 and AR 12967 and then only to AR 12967.
The connectivities from the two different MCT models are reasonably consistent though there are slight timing differences.
There is significant divergence in connectivities using the different input magnetograms after early on 2022 March 24.
In the ADAPT map SO is partially/fully connected to the positive polarity on the eastern side of the negative polarity of AR 12967 whereas in the NSO map, SO is connected to magnetic field present in the southeast direction from the positive polarity of AR 12967 and  across the solar equator.

\begin{figure}[t]
\epsscale{1.1}
\plotone{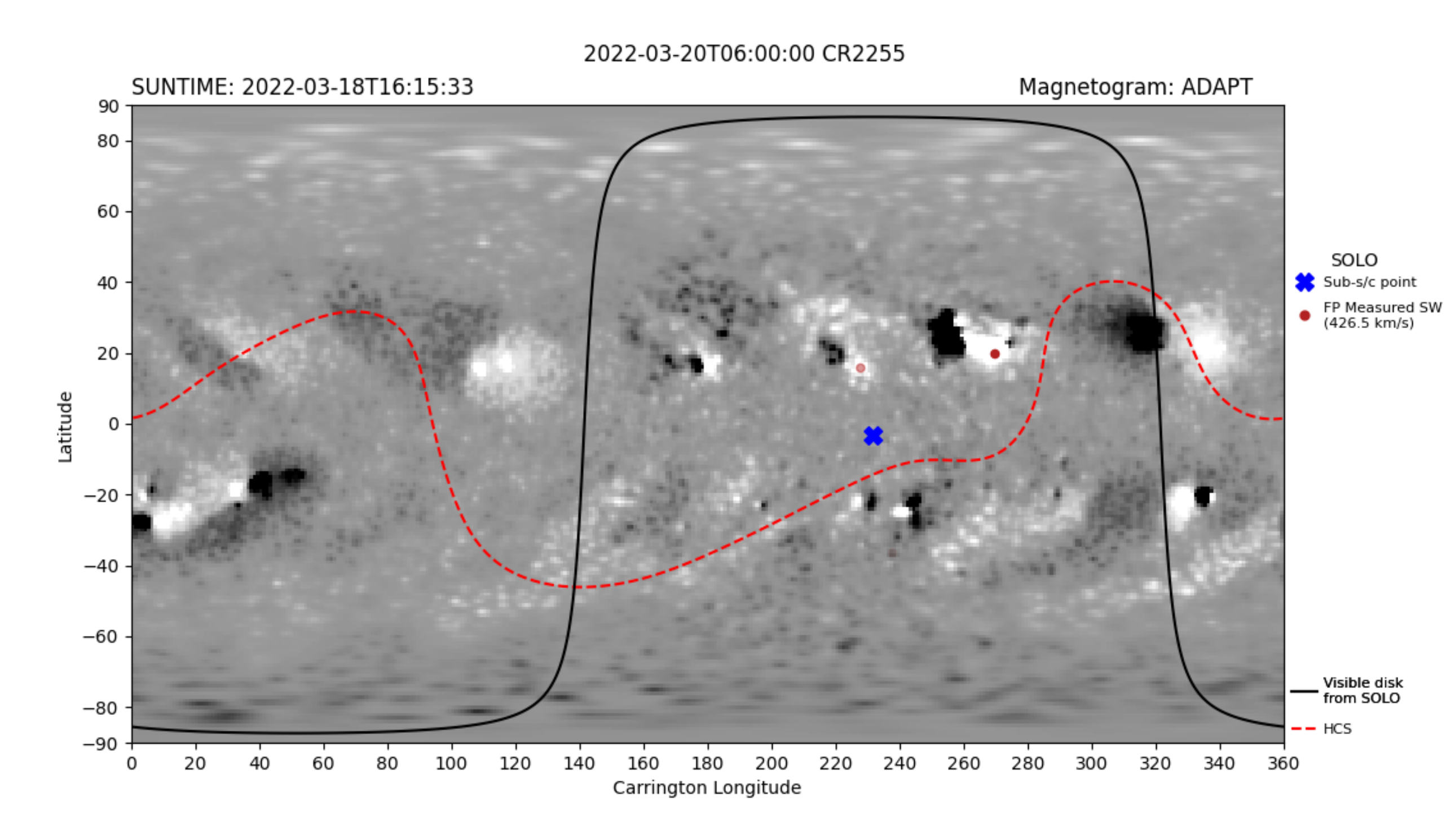}

\caption{ADAPT connectivity maps generated using the Solar Orbiter Magnetic Connectivity Tool found at  \url{http://connect-tool.irap.omp.eu/}.
Spacecraft/Sun times are 2022 March 20 06:00 UT/2022 March 18 16:15 UT.
Slow solar wind connectivities are indicated by the red dots.
The figure has an included
movie labeled MCAT$\_$ADAPT.mp4 covering the period beginning Spacecraft/Sun times of March 20 00:00 UT/2022 March 18 03:16 UT and ending March 27 00:00 UT/2022 March 25 04:58 UT (duration $\sim$2 s).
\label{fig:append1}} 
\end{figure}

\begin{figure}[t]
\epsscale{1.1}
\plotone{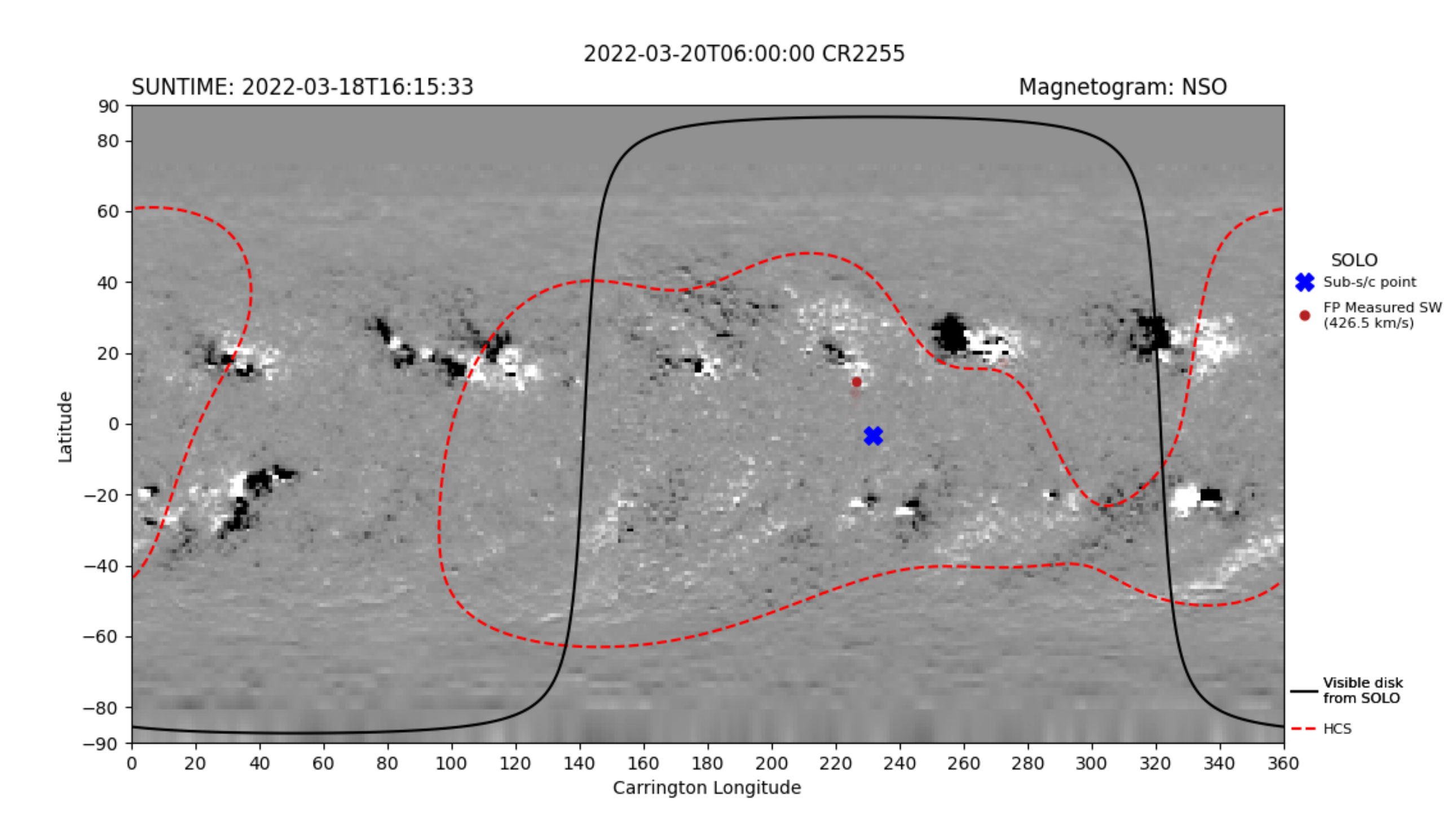}

\caption{NSO connectivity maps generated using the Solar Orbiter Magnetic Connectivity Tool found at  \url{http://connect-tool.irap.omp.eu/}.
Spacecraft/Sun times are 2022 March 20 06:00 UT/2022 March 18 16:15 UT (bottom).
Slow solar wind connectivities are indicated by the red dots.
The figure has an included
movie labeled MCAT$\_$NSO.mp4 covering the period beginning Spacecraft/Sun times of March 19 00:00 UT/2022 March 17 09:26 UT and ending March 27 00:00 UT/2022 March 25 04:58 UT (duration $\sim$2 s).
\label{fig:append2}} 
\end{figure}


\begin{acknowledgments}
We thank the reviewer for their constructive suggestions that helped to improve the manuscript.
We also thank Marc Derosa for insight into the PFSS extrapolation package used in this work.
This research used version 3.9.7 of the SunPy open source software package and the EIS Python Analysis Code (EISPAC) software package from NRL.
Solar Orbiter data were acquired from the Solar Orbiter Archive (SOAR) found at: \url{https://soar.esac.esa.int/soar/}.
Hinode is a Japanese mission developed and launched by ISAS/JAXA, collaborating with NAOJ as a domestic partner, and NASA and STFC (UK) as international partners. 
Scientific operation of Hinode is performed by the Hinode science team organized at ISAS/JAXA. 
This team mainly consists of scientists from institutes in the partner countries. 
Support for the post-launch operation is provided by JAXA and NAOJ (Japan), STFC (UK), NASA, ESA, and NSC (Norway). 
Solar Orbiter is a space mission of international collaboration between 
ESA and NASA, operated by ESA. Solar Orbiter Solar Wind Analyser (SWA) 
data are derived from scientific sensors which have been designed and 
created, and are operated under funding provided in numerous contracts 
from the UK Space Agency (UKSA), the UK Science and Technology 
Facilities Council (STFC), the Agenzia Spaziale Italiana (ASI), the 
Centre National d’Etudes Spatiales (CNES, France), the Centre National 
de la Recherche Scientifique (CNRS, France), the Czech contribution
to the ESA PRODEX programme and NASA. Solar Orbiter SWA work at UCL/MSSL 
was funded under STFC grants ST/T001356/1, ST/S000240/1, ST/X002152/1 
and ST/W001004/1.
We thank the Solar Orbiter Operations and MADAWG teams for their dedicated efforts to make this research possible.
D.B. is funded under STFC consolidated grant number ST/S000240/1.
S.L.Y. would like to thank NERC for funding via the SWIMMR Aviation Risk Modelling (SWARM) Project, grant number NE/V002899/1 and STFC via the consolidated grant number STFC ST/V000497/1.
L.v.D.G. acknowledges the Hungarian National Research, Development and Innovation Office grant OTKA K-131508.
D.M.L. is grateful to the Science Technology and Facilities Council for the award of an Ernest Rutherford Fellowship (ST/R003246/1).
The work of D.H.B. was performed under contract to the Naval Research Laboratory and was funded by the NASA Hinode program. 
A.W.J. was supported by a European Space Agency (ESA) Research Fellowship.
The ROB team thanks the Belgian Federal Science Policy Office (BELSPO) for the provision of financial support in the framework of the PRODEX Programme of the European Space Agency (ESA) under contract numbers 4000134088, 4000112292, 4000136424, and 4000134474.
We recognise the collaborative and open nature of knowledge creation and dissemination, under the control of the academic community as expressed by Camille No\^{u}s at http://www.cogitamus.fr/indexen.html.
\end{acknowledgments}

\bibliography{references}{}
\bibliographystyle{aasjournal}

\end{document}